\documentclass[twocolumn]{aastex631}
\usepackage{placeins}

\newcommand{\water}{H$_2$O}
\newcommand{\kp}{$K_p$}
\newcommand{\vrad}{$v_{\text{rad}}$}

\usepackage{array}
\usepackage{booktabs}

\begin{document}

\title{Unraveling the Mystery of the Peculiar and Young Hot Jupiter CoRoT-2b I: H$_2$O and CO Detection from Dayside Observations with Gemini-S/IGRINS}

\author[0009-0002-5701-6276]{Ying Shu}
\affiliation{Trottier Institute for Research on Exoplanets and D\'epartement de Physique, Université de Montréal, 1375 Avenue Thérèse-Lavoie-Roux, Montréal, QC, H2V 0B3, Canada}
\affiliation{Department of Physics and Astronomy, University of Waterloo, 200 University Avenue West, Waterloo, ON, N2L 3G1, Canada}

\author[0000-0003-4987-6591]{Lisa Dang}
\affiliation{Trottier Institute for Research on Exoplanets and D\'epartement de Physique, Université de Montréal, 1375 Avenue Thérèse-Lavoie-Roux, Montréal, QC, H2V 0B3, Canada}
\affiliation{Department of Physics and Astronomy, University of Waterloo, 200 University Avenue West, Waterloo, ON, N2L 3G1, Canada}

\author[0000-0002-7786-0661]{Antoine Darveau-Bernier}
\affiliation{Trottier Institute for Research on Exoplanets and D\'epartement de Physique, Université de Montréal, 1375 Avenue Thérèse-Lavoie-Roux, Montréal, QC, H2V 0B3, Canada}

\author[0000-0002-3239-5989]{Aurora Kesseli}
\affiliation{IPAC, Mail Code 100-22, Caltech, 1200 E. California Blvd., Pasadena, CA 91125, USA}

\author[0000-0003-3181-5264]{Luc Bazinet}
\affiliation{Trottier Institute for Research on Exoplanets and D\'epartement de Physique, Université de Montréal, 1375 Avenue Thérèse-Lavoie-Roux, Montréal, QC, H2V 0B3, Canada}

\author{Justin Lipper}
\affiliation{Department of Physics, McGill University, 3600 University St, Montreal, QC H3A 2T8, Canada}
\affiliation{Trottier Institute for Research on Exoplanets and D\'epartement de Physique, Université de Montréal, 1375 Avenue Thérèse-Lavoie-Roux, Montréal, QC, H2V 0B3, Canada}

\author[0000-0002-8573-805X]{Stefan Pelletier}
\affiliation{Observatoire astronomique de l’Universit\'e de Gen\`eve, 51 chemin Pegasi 1290 Versoix, Switzerland}

\author[0000-0002-1199-9759]{Romain Allart}
\affiliation{Trottier Institute for Research on Exoplanets and D\'epartement de Physique, Université de Montréal, 1375 Avenue Thérèse-Lavoie-Roux, Montréal, QC, H2V 0B3, Canada}

\author[0000-0001-6129-5699]{Nicolas B.\ Cowan}
\affiliation{Department of Physics, McGill University, 3600 University St, Montreal, QC H3A 2T8, Canada}
\affiliation{Department of Earth \& Planetary Sciences, McGill University, 3450 University St, Montréal, H3A 2A7, Canada}

\author[0000-0002-6780-4252]{David Lafreni\'ere}
\affiliation{Trottier Institute for Research on Exoplanets and D\'epartement de Physique, Université de Montréal, 1375 Avenue Thérèse-Lavoie-Roux, Montréal, QC, H2V 0B3, Canada}

\author[0000-0003-3963-9672]{Emily Rauscher}
\affiliation{Department of Astronomy, University of Michigan, 1085 S. University, Ann Arbor, MI 48109}

\author[0000-0002-0516-7956]{Alejandro S\'anchez-L\'opez}
\affiliation{Instituto de Astrofísica de Andalucía (IAA-CSIC), Gta. de la Astronomía, s/n, Genil, E-18008 Granada, Spain}

\author[0000-0001-5578-1498]{Bj\"orn Benneke}
\affiliation{Trottier Institute for Research on Exoplanets and D\'epartement de Physique, Université de Montréal, 1375 Avenue Thérèse-Lavoie-Roux, Montréal, QC, H2V 0B3, Canada}

\author[0000-0001-9427-1642]{Anne Boucher}
\affiliation{Department of Physics, McGill University, 3600 University St, Montreal, QC H3A 2T8, Canada}
\affiliation{Trottier Institute for Research on Exoplanets and D\'epartement de Physique, Université de Montréal, 1375 Avenue Thérèse-Lavoie-Roux, Montréal, QC, H2V 0B3, Canada}

\author[0000-0001-5485-4675]{Ren\'e Doyon}
\affiliation{Trottier Institute for Research on Exoplanets and D\'epartement de Physique, Université de Montréal, 1375 Avenue Thérèse-Lavoie-Roux, Montréal, QC, H2V 0B3, Canada}

\begin{abstract}

We present ground-based high-resolution spectroscopic pre-eclipse observations of the hot Jupiter CoRoT-2b obtained with the IGRINS spectrograph on Gemini South. Using cross-correlation analysis, we detect the Doppler-shifted signature of the planet's thermal emission with a signal-to-noise ratio of 4.32. Our independent analyses confirm the presence of \water{} with a confidence level of 2.6$\sigma$ and an abundance of log$_{10}$$-5.08^{+0.43}_{-0.43}$, as well as CO with 2.3$\sigma$ confidence and an abundance of
log$_{10}$$-4.21^{+0.48}_{-0.81}$ in CoRoT-2b's atmosphere, using two fully independent data reduction and retrieval pipelines. No significant detections of CH$_4$, CO$_2$, TiO, or VO are reported. While our cross-correlation analysis tentatively suggests the presence of HCN and OH, retrieval analysis does not confirm these molecules. The detected \water{} and CO features indicate that CoRoT-2b's dayside spectrum is not featureless, as previously inferred from lower-resolution observations, but instead reveals a complex atmospheric structure. Interestingly, we find a lack of significant molecular features at wavelengths shorter than 1.7 $\mu m$, potentially due to high-altitude absorbers such as H$^-$, clouds, or observational systematics. From our retrieved abundances of CO and \water{}, we constrain a supersolar C/O ratio of $0.91^{+0.08}_{-0.17}$ and a subsolar metallicity. This study provides the first high-resolution constraints on the atmospheric composition of CoRoT-2b and serves as the foundation for future investigations into its peculiar westward hotspot offset. Further phase-resolved observations will be required to explore the underlying atmospheric dynamics in more detail.
\end{abstract}


\section{Introduction} \label{sec:intro}

Measuring the chemical compositions and thermal structures of gas giants can yield valuable insights not only into the physical and chemical processes that govern their atmospheres, but also into potential formation pathways as their primordial atmosphere is a record of their evolutionary history \citep{madhusudhan_overview_2019}.
With no analog in the Solar System, hot Jupiters are gas giant planets that orbit their parent stars with orbital periods of less than a few days \citep{fortney_hotJup_2021}. They are the optimal exoplanets for atmospheric detection from emission spectroscopy due to their elevated temperatures and close proximity to their host stars \citep{seager_atmosphere_2010}. With semi-major axes less than about 0.05\,AU, hot Jupiters are expected to be synchronously rotating due to tidal interaction with their host stars, resulting in a permanent dayside and nightside. Blasted by enormous amounts of stellar flux, they can exhibit significant day–night temperature contrasts ranging from hundreds to thousands of degrees. Given their close orbits to their host stars, the dayside of hot Jupiters can reach temperatures well upwards of 1000 K \citep{fortney_hotJup_2021}.

 While space-based instruments with lower spectral resolutions aboard HST, Spitzer and James Webb Space Telescope have successfully detected molecular broad band features in transit and eclipse spectra of exoplanets, ground-based High Resolution (R=$\lambda / \Delta \lambda$$>$20000) Cross-Correlation Spectroscopy (HRCCS) has opened new avenues for exoplanetary spectroscopy by enabling unique constraints in relative chemical abundances and dynamics by resolving individual spectral lines. Since the radial velocity (RV) signal of the planet is orders of magnitude greater than that of the star, HRCCS leverages the planet's large Doppler shifts to disentangle the time-varying planetary lines from stationary or quasi-stationary telluric and stellar features \citep{brogi_highres_2021}. 

At optical wavelengths, the brightness of a star often dominates, therefore diminishing the relative amount of flux from the planet. However, the thermal emission of hot Jupiters peaks in the infrared allowing the relative flux from the planet to become more prominent. Consequently, the flux from the planet is dominated by thermal emission rather than reflection, making the planet bright enough to allow high precision measurements \citep{seager_atmosphere_2010}. Due to the higher planet-to-star contrast ratio at longer wavelength, infrared spectroscopy emerges as a more favorable tool for exploring  exoplanetary atmospheres. This approach enables simultaneous inference of both the thermal profile and vertical distribution of atmospheric species \citep{encrenaz_infrared_2014}. 

Recent advancements in ground-based high-resolution spectroscopy have enabled precise examination of the composition of exoplanetary atmospheres. The use of instruments such as IGRINS/GeminiS \citep[R=45,000;][]{line_wasp77_2021, brogi_pastobs_2023}, CFHT/SPIRou \citep[R=70,000;][]{pelletier_where_2021, boucher_co_2023, allart_pastobs_2023}, VLT/CRIRES+ \citep[R=100,000;][]{yan_crires_2023, landman_pastobs_2024}, NIRPS \citep[R=80,000;][]{bouchy_2017, artigau_2024}, CARMENES \citep[R=80,000;][]{czesla_2024, Alonso_2019} and GIANO \citep[R=50,000;][]{oliva_2018, Giacobbe_2021, carleo_2022, Guilluy_2022}, have enabled successful detections of a diverse range of chemical species \citep{borsa_2021}. 

The CoRoT-2 system stands out among the multitude of known hot Jupiters due to three distinctive features: its relatively young and active host star, the planet's unusually inflated radius given its mass \citep{guillot_c_age_2011}, and its featureless emission spectrum \citep{wilkins_pastobs_2014, Dang_c2b_2018}. CoRoT-2b is a transiting hot Jupiter \citep{alonso_pastobs_2008}, discovered during the first $\sim$150-d pointing of the CoRoT mission \citep{baglin_c2b_2006}. It has 3.47 Jupiter masses and an inflated radius of 1.466 $R_{\text{jup}}$, orbiting a relatively young G dwarf star (G7V, $T_{\text{eff}} = 5625$ K) with an orbital period of 1.743 days (see Table \ref{tab:parameters} for the stellar and planetary parameters for the system). The age of its host star, CoRoT-2, was estimated to be below 500 Ma with some models suggesting that it could be as young as 30 Ma \citep{guillot_c_age_2011}. CoRoT-2b has been the target of numerous observations, including William Herschel Telescope/LIRIS \citep{alonso_pastobs_2010}, Spitzer Space Telescope \citep{Deming_pastobs_2011, Dang_c2b_2018}, and Hubble/WFC3 \citep{wilkins_pastobs_2014}. \cite{Moses_2013} found that disequilibrium models with C/O$\sim$1 would explain CoRoT-2b’s \textit{Spitzer} dayside emission spectrum, however, subsequent eclipse observations suggested that it is best
described by a blackbody, with no molecular features confidently detected \citep{Deming_pastobs_2011, wilkins_pastobs_2014, Dang_c2b_2018}.

\begin{table}[t]
    \caption{CoRoT-2 System Parameters}
    \centering
    \begin{tabular}{lc}
         \toprule
         \toprule
         \multicolumn{1}{l}{Stellar Parameters} & \multicolumn{1}{c}{Value} \\
         \addlinespace[2pt]
         \midrule
         Spectral type$^a$  & G7V \\
         Temperature$^b$ ($T_{\text{eff}}$) & 5625$\pm$120 K \\
         Stellar radius$^b$ ($R_\star$) & 0.902$\pm$0.018 $R_{\odot}$ \\
         Stellar mass$^b$ ($M_\star$) & 0.970$\pm$0.060 $M_{\odot}$\\
         Systemic RV$^a$ ($v_{\text{sys}}$)  & 23.245$\pm$0.010 km s$^{-1}$ \\
         \midrule
         \multicolumn{1}{l}{Planet Parameters} & \multicolumn{1}{c}{Value} \\
         \addlinespace[2pt]
         \midrule
         Planet radius$^c$ ($R_P$) & 1.466$^{+0.042}_{-0.044}$ $R_{\text{Jup}}$ \\
         Planet mass$^c$ ($M_P$)  & 3.47$\pm$0.22 $M_{\text{Jup}}$ \\
         Semi-major axis$^c$ ($a$) & 0.02798$^{+0.00076}_{-0.00080}$ AU \\
         Eccentricity$^c$ (e) & 0.0143$^{+0.0077}_{-0.0076}$ \\
         Transit time$^d$ (t$_0$) & 2457347.04314(12) days\\
         Orbital period$^d$ ($P$) & 1.742997(15) days \\
         Transit duration$^e$ ($T_{14}$) & 2.273$\pm$ 0.017 hours\\
         Equilibrium temperature$^b$ ($T_{eq}$) & 1535$\pm$21 K \\
         Planet RV semi-amplitude ($K_P$) & 174.64$\pm$4.99 km s$^{-1}$ \\
         \bottomrule
    \end{tabular}
    \label{tab:parameters}

    \medskip
    
    \begin{minipage}{\linewidth}
        \textbf{References --} $^a$\cite{alonso_pastobs_2008}; $^b$\cite{ozturk_c2b_2019}; $^c$\cite{guillon_pparameter_2010};
        $^d$\cite{kokori_c2b_2022};
        $^e$\cite{Baluev_2015}. $K_P$ is calculated by assuming a circular orbit with the reported semi-major axis and period: $K_P=2\pi a/P$, and the uncertainty is propagated using $\partial K_P = K_P[(\frac{\partial a}{a})^2+(\frac{\partial P}{P})^2]^{\frac{1}{2}}$.
   
    \end{minipage}
    
\end{table}

Over the past decade, phase-resolved observations have become instrumental in deciphering the longitudinal inhomogeneity in short-period exoplanet atmospheres. In particular, photometric full orbit phase curves with space telescopes have measured the horizontal thermal and cloud distribution in the hot Jupiter atmosphere, providing valuable insight into global atmospheric circulation on hot Jupiters \citep[e.g.][]{bell_spitzer_2021, may_exoplanet_2022, Keating_2019, Dang_2024, wong_2021,jones_2022}. \cite{Dang_c2b_2018} revealed CoRoT-2b to be an oddball: contrary to the typical phase curves of most hot Jupiters \citep{showman_2002, knutson_2007, bell_spitzer_2021, Dang_2024}, the full-orbit Spitzer broadband 4.5\,$\mu m$ phase curve of CoRoT-2b exhibits a westward hotspot offset of 23$\pm$4 degrees. They also showed that this offset is robust against different reduction and analysis schemes with Spitzer. The authors proposed three scenarios to explain the westward offset: westward equatorial winds caused by asynchronous rotation \citep{rauscher_2014}, magnetic effects on the planet’s atmosphere \citep{hindle_2021, hindle_2019}, or partial cloud coverage \citep{teinturier_2024, roman_2021}.

Synergistically with photometric phase curves, spectroscopic observations at different orbital phases offer avenues to obtain constraints on the vertical temperature structure (i.e., temperature-pressure profile) and atmospheric composition at different longitudes, enabling us to distinguish whether the phase curve variations are modulated by changes in thermal structure or in chemical composition \citep{brogi_pastobs_2023, smith_pastobs_2024, pino_2022, wardenier_2024, van_sluijs_2023, ejrenreich_2020}.

To investigate the peculiar emission of CoRoT-2b, we obtained phase-resolved ground-based observations using both IGRINS and CRIRES+. In this first paper, we present the analysis of pre-eclipse high-resolution spectra taken with the Gemini-S/IGRINS instrument (R=45,000; described in section \ref{sec:obs}). Such observations could offer an independent investigation to determine whether CoRoT-2b exhibits a steady-state westward hotspot. Previous detections of chemical species such as \water, CO, and OH from other planets with IGRINS have paved promising pathways for further exploration of CoRoT-2b's atmosphere \citep{line_wasp77_2021, brogi_pastobs_2023, smith_pastobs_2024, bazinet_highres_2024, WeinerMansfield2024}.
 
The manuscript is organized as follows: Section \ref{sec:obs} describes the observation and data reduction procedures. In Section \ref{sec:result}, we present the detection results and atmospheric retrievals. Section \ref{sec:discussion} discusses the implications of our results and a summary and conclusion is presented in section \ref{sec:conclusion}.

\section{Observations and Data Reduction} \label{sec:obs}

CoRoT-2b was observed using the Immersion GRating INfrared Spectrometer \citep[IGRINS;][]{chan_igrins_2014, gregory_igpipe_2018}, a visiting instrument at the Gemini South Observatory on Cerro Pachon, Chile, as part of observing program GS-2023A-Q-111 (PI: Dang). IGRINS is a cross-dispersed NIR spectrograph that offers simultaneous coverage of 54 spectral orders from 1.45 to 2.45 $\mu m$ ($H$ and $K$ band) in the infrared with a spectral resolution of $R\approx 45000$.

\begin{figure}[!t]
    \centering
    \includegraphics[width=0.90\linewidth]{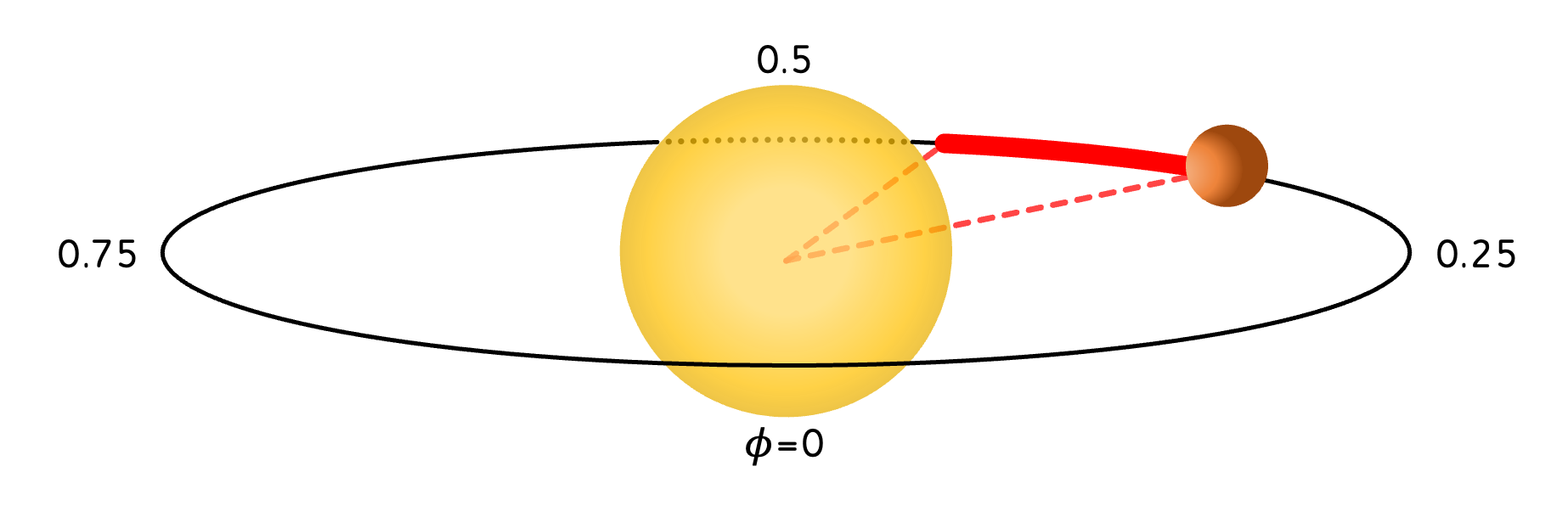}
    \includegraphics[width=1\linewidth]{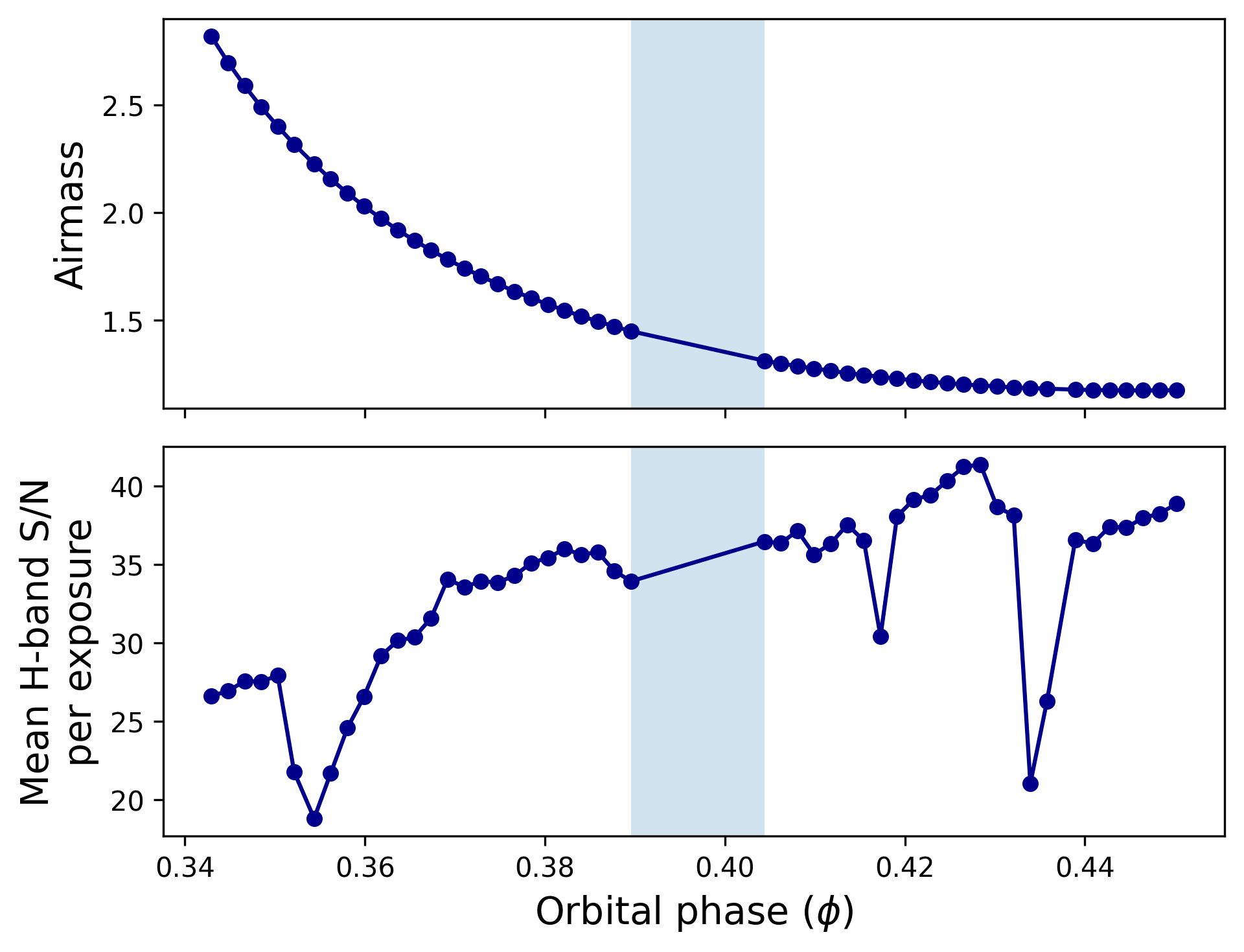}
    \includegraphics[width=1\linewidth]{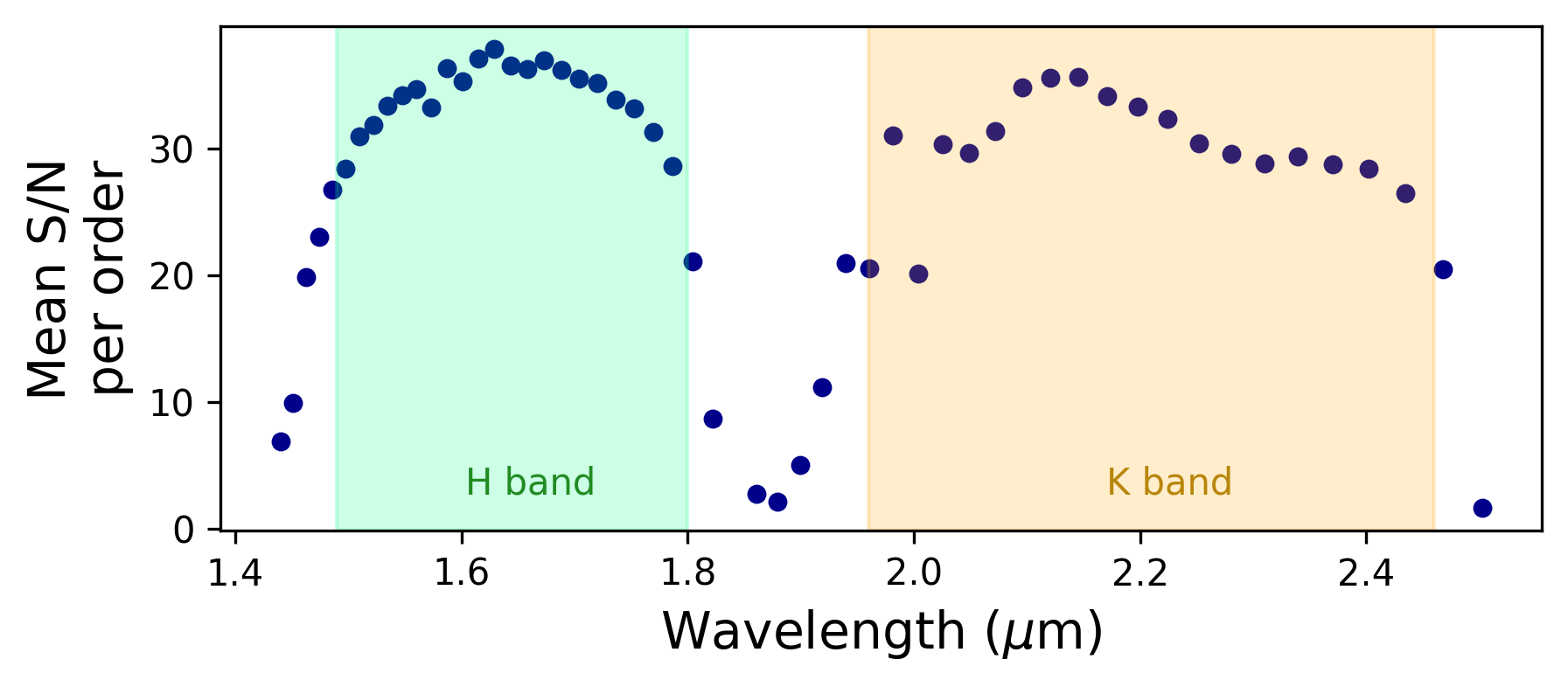}
    \caption{
    \textit{Top panel}: Depiction of the observed orbital phase, covering a phase of 0.34  $< \phi < $ 0.45. \textit{Second panel}: Airmass variation during the pre-eclipse observation. No data was collected between phases 0.390 and 0.404, highlighted in blue. Airmass conditions steadily improved throughout the observed phase. \textit{Third panel}: Time evolution of the spectral mean of the S/N in the H-band per exposure. \textit{Bottom panel}: Temporal mean of the S/N per order. The H (1.49 -- 1.80 $\mu m$) and K (1.96 -- 2.46 $\mu m$) bands are shaded in green and orange, respectively.}
    \label{fig:orbit-airmass}
\end{figure}

The data were obtained during a single continuous observation sequence (Figure \ref{fig:orbit-airmass} top panel) of the pre-eclipse phases (0.34  $< \phi < $ 0.45, where $\phi$ = 0 is transit and $\phi$ = 0.5 is secondary eclipse) on May 2nd, 2023. The observation spanned 4 hours 29 minutes (05:11 - 09:40 UT) with a total of 51 exposures. As depicted in Figure \ref{fig:orbit-airmass}, the decreasing airmass reflects the sky displacement of the star from the horizon to the zenith over the observation period. The gap between phases 0.390 and 0.404 indicates that no data were collected during this interval due to high winds impacting guiding.

The raw spectra were extracted using the IGRINS Pipeline Package \citep{lee_igpipe_2016, gregory_igpipe_2018}. Within the pipeline framework, the spectra underwent wavelength calibration and flat-field correction. To detrend the raw spectral sequence, we use \texttt{STARSHIPS} (Spectral Transmission and Radiation Search for High-resolution Planetary Signals; \citep{boucher_co_2023}), an open-source pipeline developed for processing and analysis of spectroscopic data, including data reduction, detrending, and spectral retrieval \citep{mraz_2024}. A comprehensive description of the methodology and its implementation is provided below.

\subsection{Cleaning up the spectral matrix}
Since \texttt{STARSHIPS} requires blaze-normalized input spectra, we first processed the data cube by removing the first-order continuum instrumental smooth variation in both the $H$ and $K$ bands. To estimate the blaze function of each spectral order, we first interpolated across NaN values and masked pixels to avoid gaps, and then applied a prominence threshold in \texttt{find\_peaks} function from \texttt{scipy} \citep{virtanen_scipy_2020} to ensure that only the highest local maxima (corresponding to continuum points rather than shallow telluric features) are selected. In regions strongly affected by telluric absorption, the blaze was instead constrained by neighboring orders of the same detector (H or K band), under the assumption that orders sharing the same optical path exhibit an identical blaze shape. Then, a polynomial fit was applied to the extracted continuum spectrum. In addition, for each detector, i.e. to each $H$ and $K$ band, we identified the highest signal-to-noise (SNR) orders that share the same blaze shape to produce an averaged blaze function. This ensemble-averaged blaze function is referred to as the best blaze function and serves as a replacement for the original distorted blaze functions in the bad telluric-dominated orders, since the blaze function of each order should be similar in theory. Each spectral order is then divided by its respective estimated blaze function resulting in the example spectrum in Figure~\ref{fig:reduc_example}B.

A priori, \texttt{STARSHIPS} corrects for the tellurics and stellar spectra with a Principle Components Analysis (PCA), however, we find better performance when the deepest tellurics and their wings are preemptively masked. Since there were no reference star observed during the night of our observations, we used the IGRINS observations of an A0V-type standard reference star used in \cite{line_wasp77_2021}, that we blaze-correct using the same procedure as described above, and a Vega stellar model provided by the IGRINS team \citep{gregory_igpipe_2018} to obtain a rough estimate of telluric absorption features. Based on the depth of the atmospheric transmission model, we constructed a mask for regions heavily affected by telluric lines prior to performing PCA analysis. Specifically, the transmission spectrum was first smoothed using a kernel of width 51, followed by convolution with a Gaussian kernel of width 5. The deepest 40\% of spectral channels were then identified and masked, corresponding to regions most strongly affected by telluric absorption. For each masked core, the mask was extended outward until the transmission recovered to 90\% of the continuum level, thereby excluding the wings of strong telluric lines. Subsequent preprocessing steps—detailed in the next section—included bad pixel correction, 5$\sigma$ sigma-clipping, and removal of three to five principal components to mitigate time-correlated systematics.

\subsection{Constructing CoRoT-2b emission spectrum}
\begin{figure*}[t]
    \centering
    \includegraphics[width=0.65\linewidth]{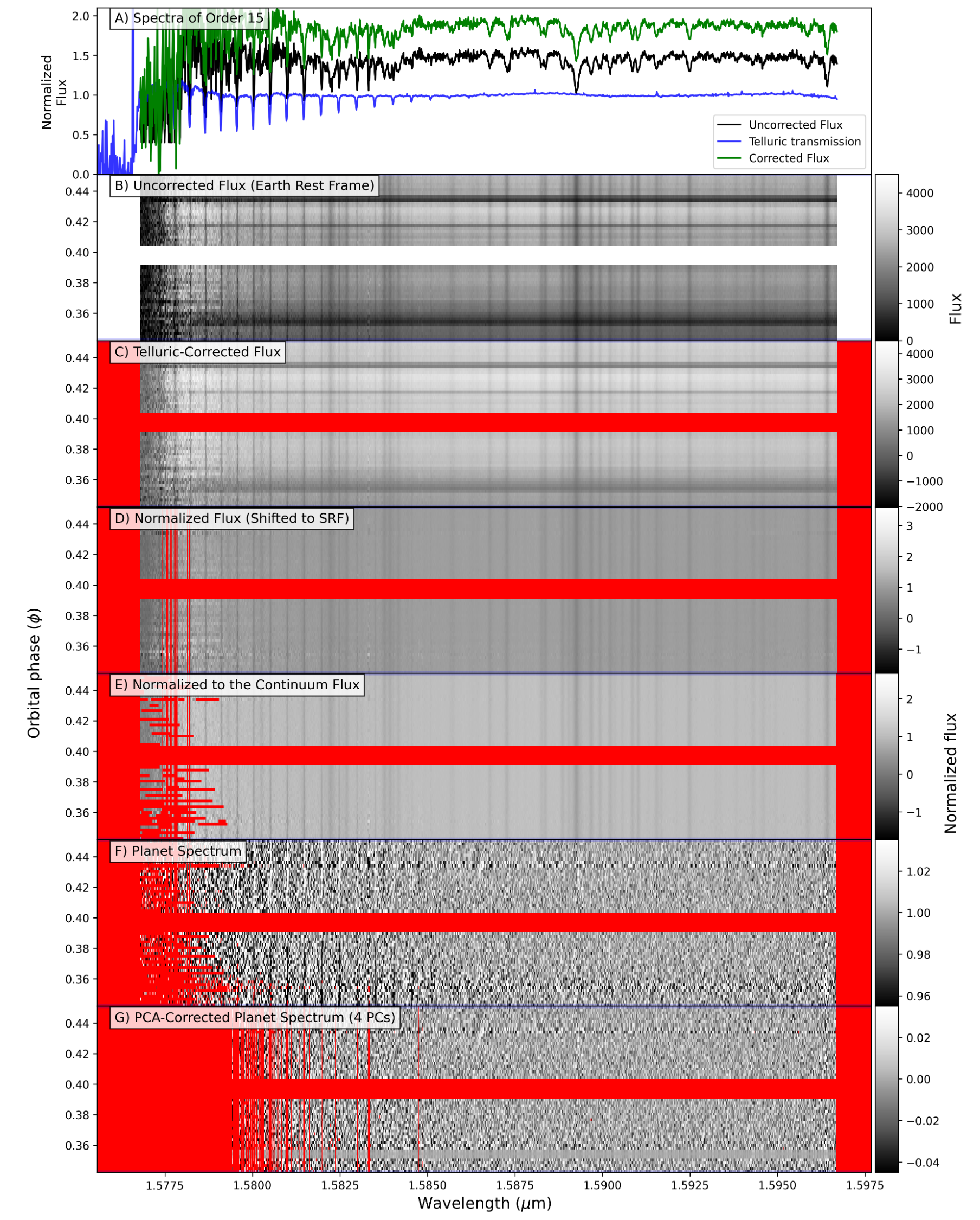}
    \caption{Example of \texttt{STARSHIPS} reduction steps that are applied to the 15\textsuperscript{th} spectral order of our IGRINS observations. The horizontal white and red bars in panels B to G are the masked spectrum due to the observation gap. \textit{Panel A}: The uncorrected (black), the telluric-corrected (green) spectra, with offsets to enhance visibility, and the reconstructed telluric transmission spectrum (blue). All three spectra have been corrected for the blaze function. The majority of the masked pixels result from imperfect removal of the blaze function. \textit{Panel B}: Uncorrected spectra. \textit{Panel C}: Telluric-corrected spectra. The masked pixels are shown in red. \textit{Panel D}: The high variance columns of the mean normalized spectra are masked, and the spectra are shifted to the pseudo star rest frame. \textit{Panel E}: The regions with deep telluric lines are masked. Every spectrum is normalized to the continuum level of the reference spectrum. \textit{Panel F}: Planetary emission spectra, where each spectrum was divided by the reference spectrum. \textit{Panel G}: Final planetary emission spectra corrected for the vertical residual structures using Principal Component Analysis (PCA) with 4 principal components. A more detailed description of each panel is provided in \cite{boucher_reduction_2021}.}
    \label{fig:reduc_example}
\end{figure*}

This step aims to remove the telluric and stellar spectra to construct the emission spectra. Figure \ref{fig:reduc_example} illustrates the complete reduction steps discussed in this section. We first correct and mask bad pixels by sigma clipping pixel (see \cite{boucher_reduction_2021}). In addition to the standard sigma clipping, due to noise dominating near the edges due to low SNR and telluric lines, resulting in a non-ideal flat spectrum, we opt to mask the initial and final 100 pixels of each order. This ensures that the corrected spectrum remains free from any continuum irregularities as shown in Figure \ref{fig:reduc_example}C. Then, we normalize each spectral order of each exposure by its median and Doppler shift each spectrum to a stellar rest frame to remove the Barycentric Earth Radial Velocity (BERV) to obtain the spectra shown in Figure \ref{fig:reduc_example}D. This allows to build a reference spectrum of the target star and the average tellurics spectrum by taking the median of all exposures together. Then, this reference spectrum is used to normalize each individual spectrum. However, some low-frequency variations in the continuum remain between exposure, which can be well characterized by a polynomial of degree 4. These polynomials are then removed from each spectra (Figure \ref{fig:reduc_example}E), allowing for a second estimate of the reference spectrum. 

Following constructing and removing the reference spectrum representative of the stellar spectrum, we divide the continuum-normalized spectra by this reference spectrum, yielding the individual planetary emission spectra with residual tellurics for each exposure as shown in Figure \ref{fig:reduc_example}F. Subsequently, we use a Principle Component Analysis (PCA) to remove any remaining quasi-static signals such as stellar and telluric residuals. In this process, we experiment with 3-5 principal components and for each degree of polynomial tested, we computed the cross-correlation function (CCF) SNR, the resulting log-likelihood and $\Delta$BIC (see Section~\ref{sec:model_comparison} and \cite{boucher_reduction_2021} Section~\ref{subsec:species in cc} for more details) to guide our choice of number of PCA components to remove. We find that the removal of 4 principal components yields the higher CCF SNR and $\Delta$BIC. Our PCA-corrected emission spectra are shown in Figure~\ref{fig:reduc_example}G.

At this stage, it is not possible to directly extract the planetary spectrum from the data, as its features are obscured by noise. However, we can benefit from the fact that the spectrum encloses a large number of spectral features to amplify their collective signal. Therefore, a forward modelling approach is first required to determine that the planetary signal is detected.

\subsection{Modeling the thermal emission spectrum for cross-correlation} \label{subsec: model_spec}

For the initial cross-correlation analysis, we first generated the planetary emission spectrum using the radiative transfer \texttt{petitRADTRANS} framework \citep[PRT;][]{molliere_prt_2019, molliere_prt_2020}. In this work, we use PRT to compute the high-resolution ($R=10^6$) emission model spectrum with 50 pressure layers between $10^{-10}$ and $10^2$ bars, log-uniformly spaced. The molecular opacities and associated line lists used include a combination of \water{}, CO \citep{Rothman_hitemp}, HCN \citep{Barber_exomol}, TiO \citep{plez_tio}, CO$_2$ \citep{Yurchenko_2020}, OH \citep{Chubb_2021, Brooke_2016}, CH$_4$ \citep{Hargreaves_2020}, and VO \citep{McKemmish_2016}. We assume constant abundances fixed to the value found at a pressure of $10^{-5}$ bar from thermochemical equilibrium. The analytical atmospheric temperature-pressure (T-P) profiles are derived from \cite{guillot_prt_2010}, who established a relation between temperature and optical depth valid for plane-parallel static grey atmospheres where we set the the mean infrared opacity to $\log \kappa = 0.1$, and the ratio of the integrated visible and infrared opacities $\log \gamma = 0.2$. We set certain parameters to specific values, namely $T_{\text{int}}$ = 500 K and $T_{\text{eq}}$ = 1693 K \citep{Dang_c2b_2018}.

After the data reduction, the planetary emission lines remain buried within the noise. We then employ the cross-correlation technique that is crucial for unveiling the planetary atmospheric signal by comparing it with the template spectra. We noted here that, during the cross-correlation process, we adopt a stellar systemic velocity of -23.245\,km\,s$^{-1}$ \citep{alonso_pastobs_2008} to ensure that the peak of the cross-correlation analysis is approximately centered at zero km\,s$^{-1}$. During this step, we elect to exclude order 30 at the central wavelength of 1.88\,$\mu m$ from the analysis owing to the unusual shape of the blaze function.

\begin{table*}[!t]
    \centering
    \caption{MCMC priors and posteriors for our 4 free-retrievals.}   
    
    \begin{tabular}{m{2cm} m{2.2cm} m{2.4cm} m{2.4cm} m{2.4cm} m{2.4cm} c}
    \toprule
    \toprule
    \multirow{2}{*}{Parameter} & \multirow{2}{*}{Prior} & \multicolumn{2}{c}{CCF Species} & \multicolumn{2}{c}{Full Species} & \multirow{2}{*}{Unit} \\
    \cmidrule(r){3-4}
    \cmidrule(r){5-6}
    & & \multicolumn{1}{c}{HRS Only} & \multicolumn{1}{c}{Joint Retrieval} & \multicolumn{1}{c}{HRS Only} & \multicolumn{1}{c}{Joint Retrieval} \\ 
    
    \addlinespace[2pt]
    \midrule

    \addlinespace[3pt]
    log$_{10}$\water & U(-12.0, -0.5) & $-4.58^{+0.80}_{-0.66}$ & $-5.07^{+0.25}_{-0.27}$ & $-4.50^{+0.72}_{-0.60}$ & $-5.08^{+0.43}_{-0.43}$ & ...\\  
    
    \addlinespace[3pt]
    log$_{10}$CO     & U(-12.0, -0.5) & $-3.97^{+0.88}_{-0.80}$ & $-4.28^{+0.40}_{-0.44}$  & $-3.97^{+0.79}_{-0.69}$ & $-4.21^{+0.48}_{-0.81}$  & ...\\
    
    \addlinespace[3pt]
    log$_{10}$HCN     & U(-12.0, -0.5) & $<-3.41$ & $<-5.92$ & $<-4.11$ & $<-5.40$  & ...\\
    
    \addlinespace[3pt]
    log$_{10}$OH     & U(-12.0, -0.5) & $<-1.71$ & $<-3.89$ & $<-0.50$ & $<-3.34$ & ... \\
    
    \addlinespace[3pt]
    log$_{10}$VO     & U(-12.0, -0.5) & ... & ... & $<-0.50$ & $<-0.50$ & ... \\
    
    \addlinespace[3pt]
    log$_{10}$TiO     & U(-12.0, -0.5) & ... & ... & $<-3.23$ & $<-4.22$ & ... \\
    
    \addlinespace[3pt]
    log$_{10}$CH$_4$     & U(-12.0, -0.5) & ... & ... & $<-4.34$ & $<-5.84$ & ... \\
    
    \addlinespace[3pt]
    log$_{10}$CO$_2$     & U(-12.0, -0.5) & ... & ... & $<-4.46$ & $<-5.59$ & ... \\
    
    \addlinespace[3pt]
    log$_{10}$SiO     & U(-12.0, -0.5) & ... & ... & $<-2.06$ & $<-3.09$ & ... \\     
    
    \addlinespace[3pt]
    log$_{10}$H-     & U(-12.0, -0.5) & $<-0.50 $ & $<-1.34$ & $<-0.50$ & $<-0.50$ & ... \\
    
    \addlinespace[3pt]
    log$_{10}$e-     & U(-12.0, -0.5) & $<-0.50 $ & $<-0.50$ & $<-0.62$ & $<-0.50$  & ... \\
    
    \addlinespace[6pt]
    log$_{10}$P$_{\text{clouds}}$     & U(-5.0, 2)   & $-0.47^{+0.71}_{-0.96}$ & $-0.27^{+0.31}_{-0.18}$  & $-0.06^{+1.39}_{-0.63}$ & $-0.19^{+0.77}_{-0.62}$ & ... \\
    
    \addlinespace[6pt]
    T$_p$            & U(400, 3500)   & $1377.70^{+1067.93}_{-477.70}$ & $1829.26^{+166.09}_{-141.88}$ & $2200.10^{+334.49}_{-964.46}$ & $1746.89^{+281.84}_{-282.53}$ & K \\
    
    \addlinespace[6pt]
    $v_{\text{rad}}$        & U(-10, 50)     & $21.3^{+2.5}_{-2.5}$ & $20.9^{+2.1}_{-1.8}$ & $22.1^{+2.4}_{-2.5}$ & $21.4^{+1.8}_{-1.7}$ & km s$^{-1}$ \\
    
    \addlinespace[6pt]
    K$_p$            & U(140, 200)    & $166^{+4}_{-5}$ & $167^{+4}_{-4}$ & $164^{+4}_{-4}$ & $166^{+3}_{-3}$ & km s$^{-1}$ \\ 
    
    \addlinespace[3pt]
    \bottomrule
     
    \end{tabular}

    \medskip
    
    \begin{minipage}{\linewidth}
    \textbf{Note.} \emph{Detected Species}: high-resolution observations only and combined low- and high-resolution observations considering only the species that were detected with our cross-correlation analysis \emph{Full Species:} high-resolution observations only and combined low- and high-resolution observations including all detected species and other expected molecules such as CO$_2$, CH$_4$, TiO, VO and SiO. This table reports constraints or 2-$\sigma$ upper limits on species abundances, cloud pressure, planet's temperature and orbital parameters.
    
    \vspace{0.5em}
\end{minipage}
    
\label{tab:retrival_table}
\end{table*}

\subsection{Model Comparison with High-resolution Data}
\label{sec:model_comparison}
As detailed in \citep{boucher_reduction_2021}, the model-to-data comparison takes into account the impact of data reduction by first generating the planet-to-star flux ratio at each phase for a given a $K_p$ and $v_{sys}$ and then by applying the same PCA analysis. Consequently, the model varies slightly for each exposure due to the effect of the PCA process.

Based on the equations in \cite{gibson_equation_2020} and \cite{boucher_co_2023}, we calculate the weighted cross-correlation function (CCF) for every order of every spectrum:
\begin{equation}\label{eq:CCF}
    \text{CCF}(\theta,\nu_P) = \sum_{i=1}^{N} \frac{f_i \cdot m_i(\theta,\nu_P)}{\sigma^2_i},
\end{equation}
where $f_i$ represents the data spectra along with their associated uncertainties $\sigma_i$. The estimation of the noise, $\sigma_i$, is determined following the same procedure as described in \citep{boucher_co_2023}. The model $m_i$ from Section \ref{subsec: model_spec} depends on the parameter vector $\theta$, which encompasses atmospheric model parameters such as the planet's temperature $T_p$, the mean infrared opacity \citep[see][]{guillot_prt_2010} $\kappa$, and the ratio of the integrated visible and infrared opacities $\gamma$.
The parameter vector is generated for a specific orbital solution 
$\nu_P = [K_p, V_{sys}]$, where $K_p$ is the orbital radial velocity semi-amplitude and $V_{sys}$ is the systemic velocity. The index $i$ runs over all times and wavelengths in the dataset, and is summed over $N$ numbers of unmasked pixels.

Then, we map the equation \ref{eq:CCF} to a likelihood function $\ln \mathcal{L_{\text{HR}}}$, as presented in \cite{boucher_co_2023}, that can be written in a compact form:
\begin{equation}\label{eq:lnL}
    \ln \mathcal{L_{\text{HR}}} = -\frac{N}{2} \ln \frac{\chi ^2}{N}.
\end{equation}

\noindent where $\chi ^2 = \Sigma ^N _i=1 (f_i - \alpha m_i)^2/\sigma ^2 _i$ and $\alpha$ is fixed to 1. Note that, no distinction is made between spectral orders; the $\ln \mathcal{L}$ includes all orders across all exposures. 

\subsection{Atmospheric retrievals} 

To constrain the molecular abundances and the thermal structure properties of CoRoT-2b, we perform a free-retrieval wherein the abundance of individual species are assumed to be constant-in-altitude and allowed to vary independently from one another.  This is in contrast to chemical equilibrium retrievals that assume all abundance profiles follow equilibrium chemistry predictions, provided an overall metallicity and carbon-to-oxygen ratio. We perform a retrieval on the IGRINS observations only and also by combining our IGRINS observations with existing eclipse spectra obtained with HST \citep{wilkins_pastobs_2014} and Spitzer \citep{Deming_pastobs_2011, Dang_c2b_2018} and are listed in the Appendix \ref{sec: corner plots} (Table \ref{tab:low_res_points}). We use the open-source \texttt{emcee} package \citep{foreman_emcee_2013} to perform a Markov Chain Monte Carlo (MCMC) fit simultaneously to the low-resolution and high-resolution observations by maximizing the following log-likelihood function $\ln (\mathcal{L_{\text{HR}}}+\mathcal{L_{\text{LR}}})$  where $\mathcal{L_{\text{LR}}}$ is defined as

\begin{equation}
    \ln \mathcal{L_{\text{LR}}} = -\sum \frac{(f_i-m_i)^2}{\sigma_i^2}.
\end{equation}

The modeled and observed eclipse depths in the spectral bin $i$ are given by $m_i$ and $f_i$ respectively where 
\begin{equation}\label{}
    m_i=\left( \frac{R_p}{R_\star} \right)^2 \left[ \frac{F_p}{F_\star} \right]_i
\end{equation}
\noindent with $R_p/R_\star$ as the ratio of planet-to-stellar radii and $F_p/F_\star$ is the flux ratio.

Table \ref{tab:retrival_table} lists the free parameters and their corresponding prior distributions. We adopt uniform priors for all parameters. The MCMC simulation was executed with around 600 burn-in steps, 6000 sampling steps, and 160 walkers assigned to each free parameter. The walkers for \kp{} and \vrad{} are initialized in the range of (165, 175) and (18, 24), respectively. After 6000 steps, we looked at the evolution of log-probability of each walker to ensure that each walker chain converged to a maximum log-probability and remained there for at least 1000 steps. We also identified if any walkers were stuck in an unphysical part of the parameter space with a lower log-probability and removed these chains from our posterior to ensure that our parameter estimates were not biased.

\section{Results} \label{sec:result}

\subsection{Cross-correlation Results}

\begin{figure*}[!t]
    \centering
    \includegraphics[width=1\linewidth]{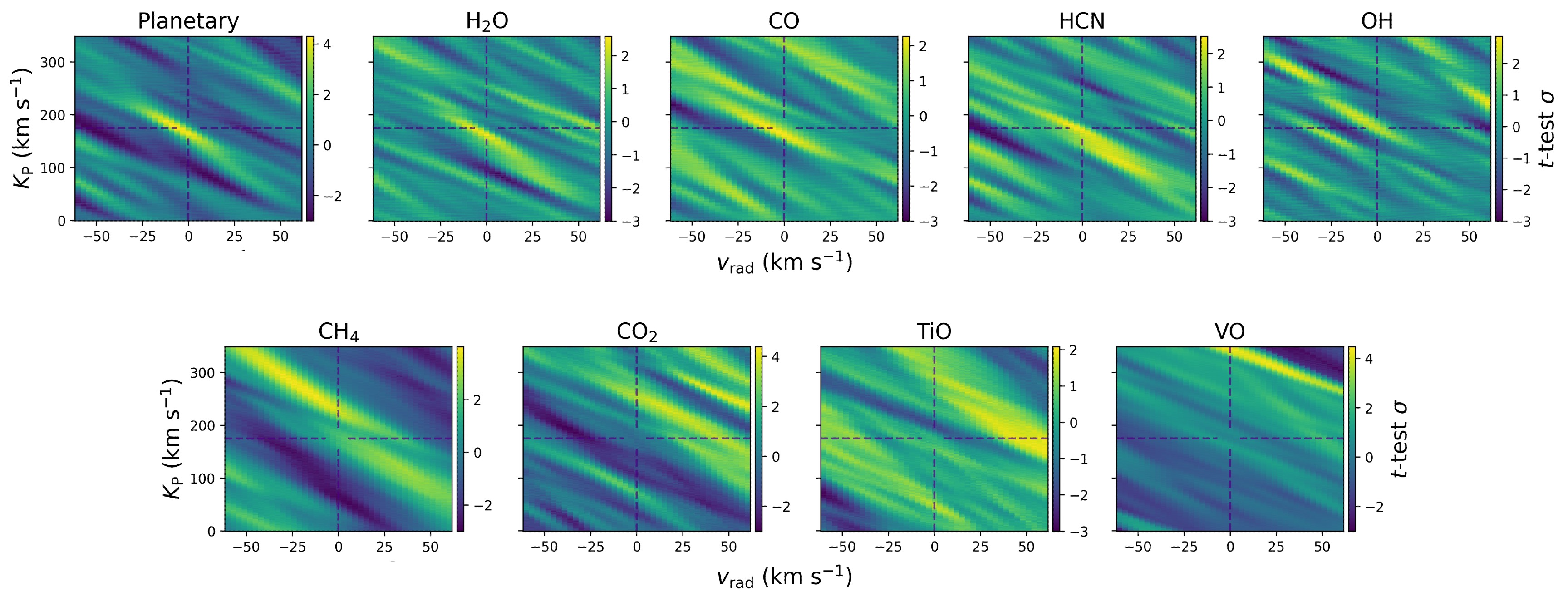}
    \caption{Cross-correlation maps of different species searched for in the atmosphere of CoRoT-2b. The detection significance of a signal is determined by computing the t-test. The dotted horizontal and vertical black lines indicate the expected \kp{} and $v_{\text{rad}}$.  Here planetary refers to the full atmospheric model. Overall, we detect H$_2$O and CO absorption, observe hints of HCN and OH, but do not find any evidence for CH$_4$, CO$_2$, TiO, or VO in the dayside atmosphere of CoRoT-2b.}
    \label{fig:comparison}
\end{figure*}

To ensure that we can detect the planetary signal, we first perform a CCF and a t-test analysis to the IGRINS CoRoT-2b observations using an atmospheric emission model including \water{} and CO generated with \texttt{petitRADTRANS}. \texttt{STARSHIPS} quantifies detection significance using a scaled Welch t-test map (see Figure~\ref{fig:comparison}), which tests the null hypothesis that two samples were drawn from the same distribution. Specifically, it compares CCF values near the planet's radial velocity to those farther away. The timestamp of each exposure is determined by calculating the mid-exposure time of the observation. These timestamps are then converted from Julian Date (JD) to Barycentric Julian Date (BJD) to ensure that the timing is corrected to the Solar System barycenter. This correction is necessary to accurately compute the radial velocity of the planet at each exposure and to properly align the spectra in the planetary rest frame during the cross-correlation analysis. To generate the detection map, CCFs are combined over time for various RV semiamplitudes (Kp) and scaled according to the corresponding t-value \citep{mraz_2024}.

\begin{figure}[!t]
    \centering
    \includegraphics[width=0.95\linewidth]{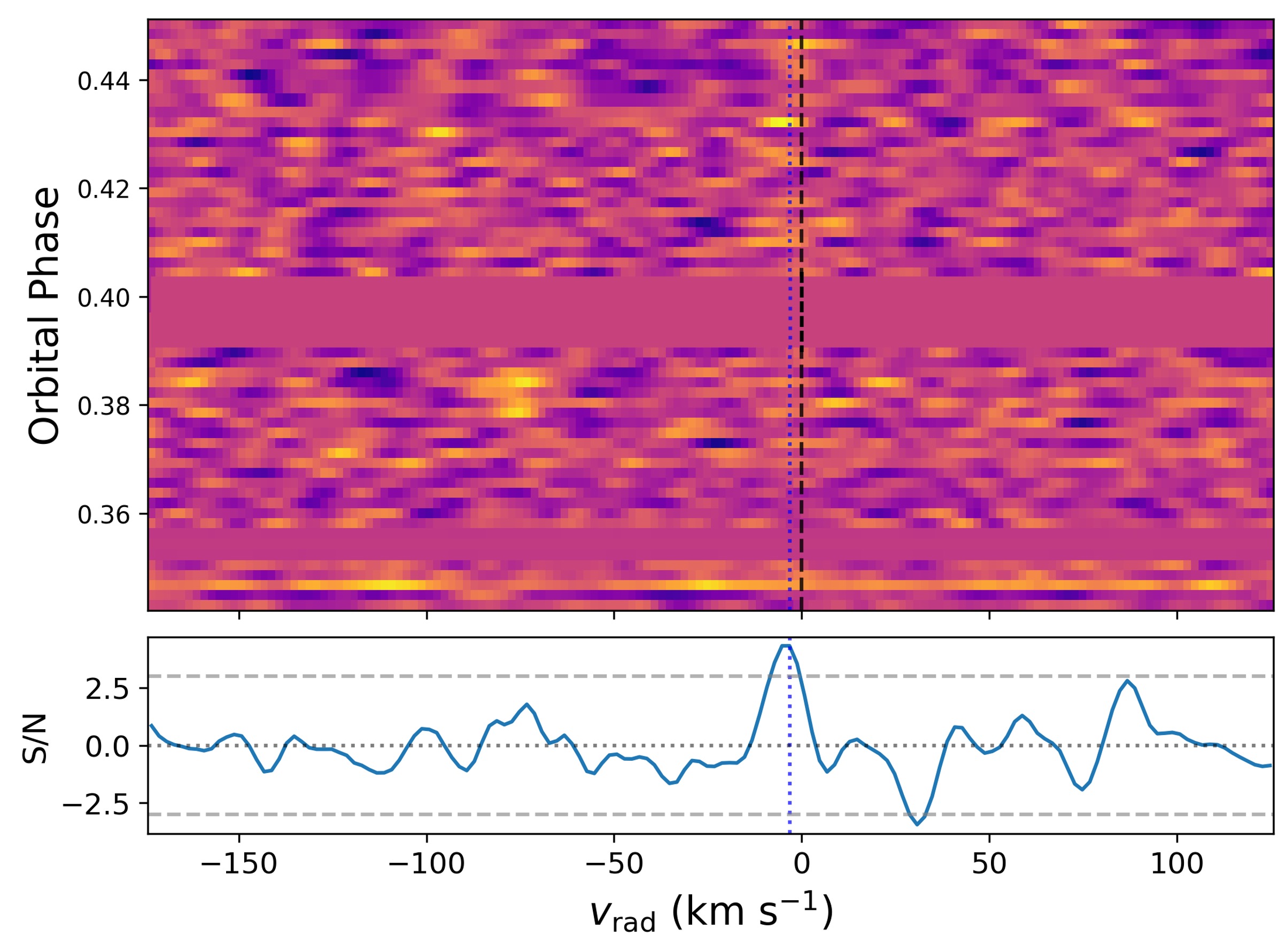}
    \caption{Planet rest-frame cross-correlation time series of the best-fit planetary model (\water{}, CO, OH, and HCN). \textit{Top panel}: Normalized CCF from individual exposures as a function of \vrad{} in the planetary rest frame by taking into account the systemic velocity of 23.245 km s$^{-1}$. The black dashed line indicates a planet path for \vrad $=0$ km s$^{-1}$. 
    \textit{Bottom panel}: 1D CCF S/N curve. The CCF is normalized by dividing by the standard deviation of the out-of-trail values (\vrad$<-25$ and $>20$ km s$^{-1}$). The planet's trace is detected at a normalized S/N of 4.32 at a \vrad{} of -3.2$\pm$1.7 km s$^{-1}$ (2$\sigma$). The grey dashed horizontal lines represent S/N of 3 and -3.
    }
    \label{fig:ccf}
\end{figure}

The CCFs of individual spectra using the best-fit planetary model are shown in Figure \ref{fig:ccf} showing a detection at a SNR of 4.32 of the planet's signal. The peak S/N corresponds to the \vrad{} of the planet at $-3.2$ km s$^{-1}$ in the planet-rest frame. Within the selected velocity range of -25 to 20 km s$^{-1}$, the corresponding 2$\sigma$ uncertainty is estimated to be $\pm$1.7 km s$^{-1}$. The t-test map comparing \kp{} to \vrad{} for the planetary, \water{} only, and CO only models is depicted in the top row of Figure \ref{fig:comparison}. Here, \kp{} represents the semi-amplitude of the planetary orbital radial velocity, while the x-axis, $v_{\text{rad}}$, denotes the RV shift relative to the planetary rest frame at a given \kp. We adjusted the RV spacing to match the Half Width Half Maximum of IGRINS and rescaled the t-test values, constraining the lower bound to -3$\sigma$ to ensure consistent statistical comparison across datasets. For pre-eclipse CoRoT-2b observation, we see \water{} with a peak t-test value of 2.6$\sigma$, CO with a peak t-test value of 2.3$\sigma$, along with tentative hints of HCN and OH. 

We only consider the signals observed for HCN and OH as being tentative as their cross-correlation maps in Figure \ref{fig:comparison} exhibit negative noise features of comparable amplitude far from the expected planetary signal.

We then generated a variety of single-molecule atmospheric models and performed a series of cross-correlations analyses to search for molecules commonly found in the atmospheres of hot Jupiters, namely \water, CO, HCN, OH, CH$_4$, CO$_2$, TiO and VO as shown in Figure \ref{fig:comparison}. Our single-molecule cross-correlation analysis shows a strong detection of \water{} and CO, as well as hints of HCN and OH absorption. In contrast, our analysis did not reveal any evidence for CO$_2$, CH$_4$, TiO, or VO (Figure \ref{fig:comparison}, bottom row).

\subsection{Abundances of Species Detected in Cross-Correlation} \label{subsec:species in cc}

\begin{figure*}[!t]
    \centering
    \includegraphics[width=\linewidth]{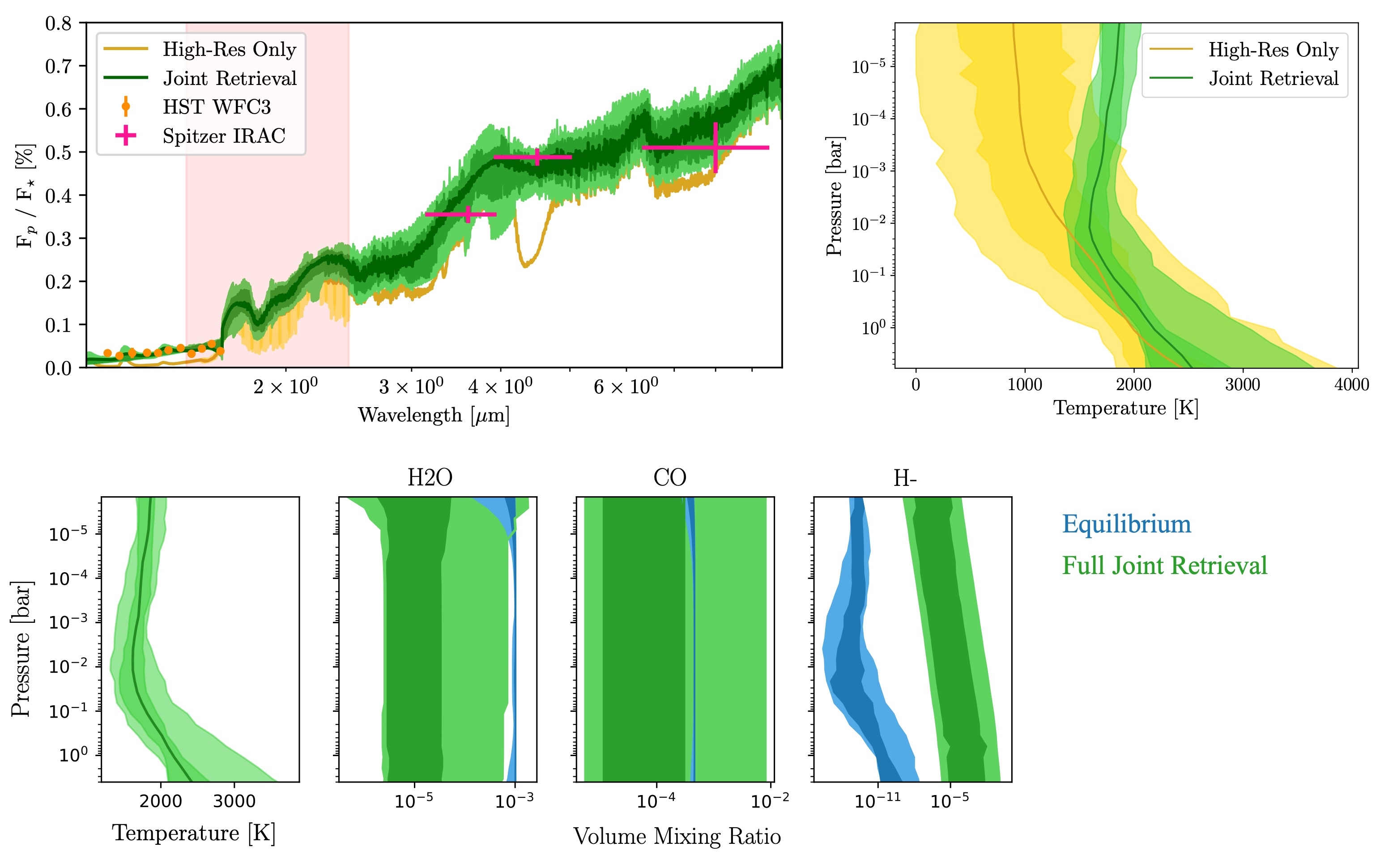}
    \caption{\emph{Top-Left:} The maximum likelihood spectra for our IGRINS-only (dark yellow) and the joint low- and high-resolution (green) retrievals with 10 atmospheric constituents (including molecules not detected in cross-correlation). We also include the 1$\sigma$ and 2-$\sigma$ confidence intervals for the joint retrieval. The shaded red region indicated the IGRINS bandpass and the HST/WFC3 and Spitzer/IRAC measurements as plotted in orange and pink, respectively. \emph{Top-Right:} The corresponding T-P profile of the high-resolution only and joint retrievals in yellow and green, respectively. The darker and lighter swaths represent the 1$\sigma$ and 2-$\sigma$ confidence intervals. \emph{Bottom:} Comparison of the abundance retrieved from our combined low- and high-resolution retrieval compared to chemical equilibrium predictions based on the retrieved vertical temperature structure.  The chemistry predictions notably do not predict any H$_2$O dissociation (i.e., OH formation) at pressures above $\sim$10\,$\mu$bar or as much H- as inferred by the free retrieval. The darker and lighter swaths represent the 1$\sigma$ and 2-$\sigma$ confidence intervals.}
    \label{fig:spectra}
\end{figure*}

To constrain the abundances of the molecules detected in cross-correlation, we ran a free-retrieval assuming only CO, \water, HCN and OH. We ran a retrieval on the IGRINS data only and another analysis combining our high-resolution observations and existing low-resolution HST/WFC3 and Spitzer eclipse spectra from \cite{wilkins_pastobs_2014} and \cite{bell_spitzer_2021}. A stellar blackbody spectrum with a temperature of 5625 K is assumed in the retrieval when fitting the data \citep{ozturk_c2b_2019}. We use log-uniform priors for all parameters and water dissociation parametrization was included for both models to allow for vertical dissociation of H$_2$O (see Appendix \ref{sec:app_water_dissociation} for details). While the dayside atmosphere of CoRoT-2b is not necessarily expected to be hot enough for water molecules to thermally dissociate, we still allow for this possibility, motivated by the observed tentative signature of OH. The priors, retrieved values of the abundances and T-P profile parameters of both fits are presented in Table \ref{tab:retrival_table}. The MCMC posterior distribution for the HRS-only and the combination of low- and high-res observations are shown in the corner plots in Appendix~\ref{sec: corner plots} (Figures \ref{fig:corner_detected} and \ref{fig:corner_joint}). We obtained a \kp{} and \vrad{} of $166^{+4}_{-5}$ km s$^{-1}$ and $21.3^{+2.5}_{-2.5}$ km s$^{-1}$ for the HRS only, which is consistent with the values of $167^{+4}_{-4}$ km s$^{-1}$ and $20.9^{+2.1}_{-1.8}$ km s$^{-1}$ when we combined all observations.

The retrievals constrain the \(\mathrm{H_2O}\) abundance to log$_{10}$ \(-4.58^{+0.80}_{-0.66}\) and the CO abundance to
log$_{10}$ \(-3.97^{+0.88}_{-0.80}\). Although the cross-correlation showed a tentative detection of HCN, the detection of HCN is not confirmed by both the HRS-only and joint retrievals as the posterior distributions do not exclude the absence of HCN. Instead, we obtain an upper limit of log$_{10}$ -3.41 (2-$\sigma$). Similarly the cross-correlation analysis showed a tentative OH detection, but both retrievals do not exhibit bounded posteriors on the abundance of OH as the absence of OH is not ruled out. The peak in OH abundances in the posterior distribution of $\log _{10} \rm{OH}=-3.77$ (HRS-only) and $\log _{10} \rm{OH}=-5.29$ (joint retrieval) seem suspiciously high relative to \(\mathrm{H_2O}\) and CO, suggesting that such abundance is unlikely. While a bounded abundance posterior may be a strict criteria for the detection of a molecule, we have also ran retrieval excluding the tentatively detected molecules, i.e HCN and OH, and computed the Akaike Information Criterion (AIC) of the best fit models with and without the molecule of interest and did not find a statistically significant improvement in AIC ($\Delta AIC < 2$) \citep{burnham2002model}.

Interestingly, although \(\mathrm{H_2O}\) is detected in our IGRINS observations, the best-fit spectra as shown in Figure~\ref{fig:spectra} from both IGRINS-only and combined with low-resolution spectra retrievals suggest muted spectral features at wavelengths below 1.7 $\mu$m. To fit this lack of feature, our free-retrieval favoring the presence of H$-$. 

\subsection{Full Retrievals Combining Low- and High-Resolution Spectra}
Since a non-detection via cross-correlation does not necessarily imply the absence of a given species in the atmosphere, we performed additional retrievals incorporating molecules such as CO$_2$, CH$_4$, TiO, VO, and SiO to place upper limits on their abundances. Moreover, adding species not detected in cross-correlation may also be necessary to set the planet continuum correctly. In addition to retrievals focused solely on molecules detected through cross-correlation analysis, we performed analogous retrievals using only high-resolution IGRINS observations and joint retrievals with existing HST/WFC3 and Spitzer data. The constraints on \water{} and CO remained consistent in all analyses, as did those for HCN and OH as listed in Table \ref{tab:retrival_table}. 

Interestingly, the full retrieval favors the presence of VO in both high-resolution and joint datasets, although our initial cross-correlation analysis was not suggestive of a VO detection. However, we note that the inferred VO abundance seems unrealistically high --up to $\sim$10000$\times$solar--, suggesting that the retrieved high abundance of VO could be acting as a filler gas to mask water features and better fit both the IGRINS and WFC3 data.  Potentially the retrieval opts to add VO to compensate for missing opacities or reflected light contribution in our retrieval. As such, we do not claim a detection of VO. 

The IGRINS-only retrieval do not exhibit a thermal inversion, as these observations are not sensitive to high altitudes of pressure ranges below 10$^{-3}$ bar; however, our joint retrieval also suggests a high-altitude thermal inversion to explain the WFC3 spectrum while allowing deeper atmospheric layers containing \water{} to be seen in absorption at longer wavelengths. Finally, abundances for TiO, CO$_2$, CH$_4$ and SiO are unconstrained for both high-resolution retrieval and joint retrieval.

\section{Discussion}\label{sec:discussion}
\begin{figure*}[t]
    \centering
    \includegraphics[width=0.96\linewidth]{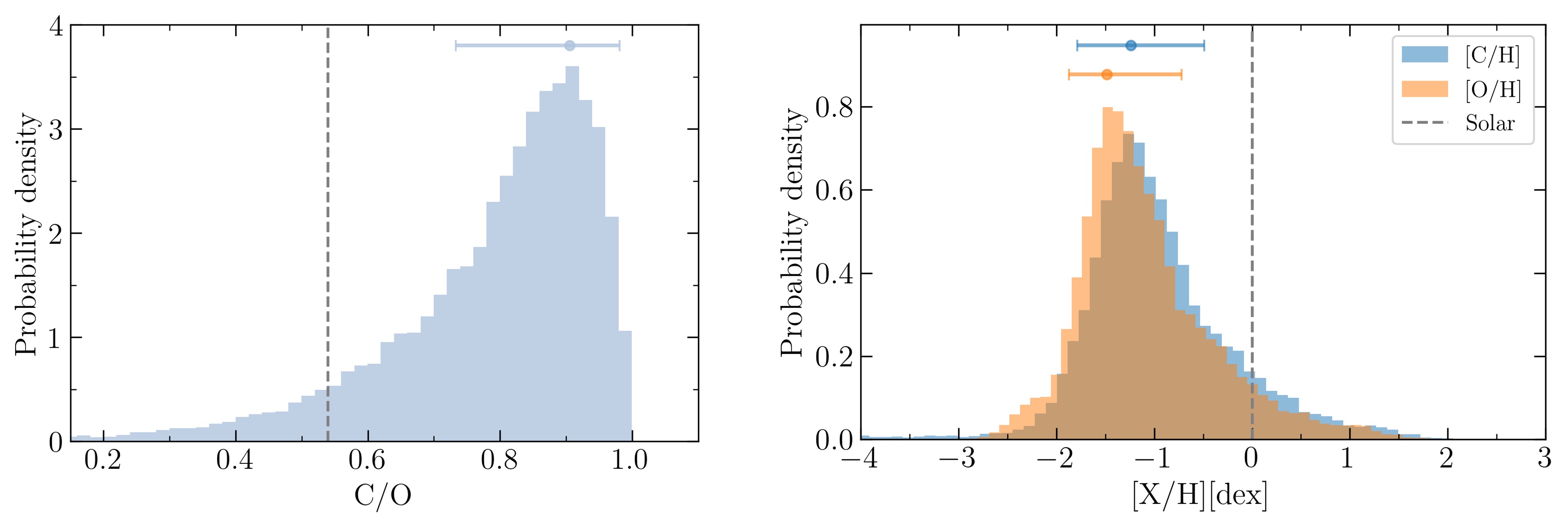}

    \caption{Elemental abundance ratios in the atmosphere of CoRoT-2b.  \emph{Left:} The blue distribution represents the C/O computed based on the MCMC posterior of the abundance of carbon- and oxygen-bearing molecules from our joint retrieval. The dashed grey line marks the solar C/O ratio for reference.
    \emph{Right:} Abundance ratio of Carbon and Oxygen relative to Hydrogen.}
    \label{fig:co_ratio}
\end{figure*}

CoRoT-2b stands out as an unusual planet for several reasons: it orbits a young active star, possesses an anomalously inflated radius, and exhibits a unique westward hotspot offset in its Spitzer phase curve \citep{guillot_c_age_2011, guillon_pparameter_2010, Dang_c2b_2018}. Previous analyses of CoRoT-2b’s dayside emission spectrum have suggested that it is best described by a blackbody, with no molecular features confidently detected \citep{Deming_pastobs_2011, wilkins_pastobs_2014, Dang_c2b_2018}. However, with our new high-spectral-resolution IGRINS observations, we report the first detection of \water{} and CO, with an abundance of log$_{10}$$-5.08^{+0.43}_{-0.43}$ and $_{10}$$-4.21^{+0.48}_{-0.81}$, further demonstrating the capability of ground-based infrared high-resolution instruments to constrain the atmospheric composition of gas giants. Moreover, our observations provide an opportunity to revisit previously reported enigmatic properties of CoRoT-2b, including its seemingly featureless dayside spectrum, the anomalously high C/O ratio, and the longitudinal distribution of its atmospheric properties.

\subsection{Lack of Features Below 1.7 \texorpdfstring{$\mu$m}{um}}
Our detection of \water{} and CO shows that the dayside spectrum of CoRoT-2b is not as featureless as previously thought. To further confirm our detections in section~\ref{sec:result}, we performed a independent reduction and retrieval of the IGRINS observations following \cite{bazinet_highres_2024} in Appendix \ref{sec:app_indep} and find abundances for \water{} and CO consistent within 1$\sigma$. However, both IGRINS and WFC3 observations are best fitted by models with a clear lack of water features at short wavelengths ($<$1.7 $\mu$m), while water absorption is confidently detected at longer wavelengths ($>$1.7 $\mu$m), 
hinting towards the presence of additional high altitude chromatic absorbers that are masking H$_2$O lines below $\sim$1.7\,$\mu$m.
Although our retrieval attempts to reconcile these observations by invoking improbably high abundances of H- and VO and a high-altitude thermal inversion, several plausible explanations could account for the absence of features in the bluer wavelengths. Given that CoRoT-2b’s dayside temperature ($\sim 1700$\,K) lies within a regime conducive to cloud formation --and that inhomogeneous dayside clouds have been proposed to explain the westward hotspot offset at 4.5 $\mu$m-- it is possible that clouds on the dayside are muting short-wavelength features \citep{Dang_c2b_2018}. Although CoRoT-2b is likely too cool for significant H$^-$ dissociation, its host star is a young (30-500 Myrs) which might lead to enhance UV radiation potentially driving H$^-$ dissociation in the upper atmosphere, thereby contributing to damping features below 1.7\,$\mu$m.

\subsection{C/O and Metallicity}

With an equilibrium temperature of $\sim$ 1700K, CoRoT-2b lies in a thermal regime where carbon and oxygen are expected to be predominantly held in gaseous CO and \water{} under thermochemical equilibrium. Assuming that no other major C- or O- bearing molecules are present, we use our retrieved abundances of CO and water obtained from our all-species joint retrieval to estimate the C/O ratio as:
\begin{equation}
\label{eq:co_ratio}
\text{C/O}_{\text{gas}} = \frac{\text{CO}}{\text{CO}+\text{H}_2\text{O}},
\end{equation}

Using our retrieved abundances of CO and \water{}, we derive a super-solar C/O ratio of $0.91^{+0.08}_{-0.17}$ (1$ \sigma$), as shown in Figure \ref{fig:co_ratio}. 
The bottom row of Figure \ref{fig:spectra} compares the retrieved abundances to volume mixing ratios predicted by a chemical equilibrium model, revealing that the elevated C/O ratio is primarily driven by the low abundance of H$_2$O relative to CO. Additionally, we derived the atmospheric metallicity of CoRoT-2b by estimating [C/H] and [O/H] of $-1.24_{-0.55}^{+0.75}$ and $-1.48_{-0.39}^{+0.76}$
(1$\sigma$), finding overall agreement with solar abundances within 2$\sigma$, though slightly sub-solar (Figure \ref{fig:co_ratio}). We note that our retrieval suggests a high H- abundance to dampen features below 1.7 $\mu$m, primarily driven by the WFC3 points.  As H- (bound-free) is not expected to be a dominant absorber below $\sim$2200\,K when molecular hydrogen begins to thermally dissociate, this is difficult to explain and could indicate that there is some missing physics or opacities in our retrieval to realistically explain the dampened features at shorter wavelengths. As a result, our retrieval could be artificially decreasing the abundance of \water{}, therefore also lowering the abundance of CO since high-resolution spectra are sensitive to relative abundances. This could lead to a bias towards lower metallicities.

The supersolar atmospheric C/O ratio combined with subsolar metallicity inferred for CoRoT-2b is consistent with trends derived from other IGRINS observations of planets in similar temperature regimes, such as WASP-77A b \citep{line_wasp77_2021, smith_pastobs_2024, bitsch_2022, khorshid_2023, coria_2024} and HIP-65 A b \citep{bazinet_highres_2024}. Assuming that CoRoT-2b’s atmospheric abundances reflect the composition of its bulk envelope, one potential formation pathway is that the planet formed beyond the major ice lines (H$_2$O, CO, CO$_2$), followed by inward migration after the dissipation of the protoplanetary disk. 
Beyond the \water{} ice line, oxygen is preferentially sequestered into solid water ice, which could enrich the gas phase in carbon relative to oxygen. A hot Jupiter accreting a substantial fraction of its envelope in this region may therefore acquire a comparatively carbon-rich and oxygen-poor atmospheric composition. The inferred low metallicity could further suggest relatively inefficient accretion of solid material, such as icy planetesimals or pebbles, both during envelope accretion and potentially throughout later migration, although alternative explanations cannot be ruled out.

\subsection{Investigating the Spatial Distribution of Species}

As our detections of H$_2$O and CO do not perfectly overlap in cross-correlation space as in other IGRINS observations of hot Jupiter in emission \citep{brogi_pastobs_2023}, we investigate this with our data by performing different retrievals allowing for different composition of the atmosphere based on the result from the cross-correlation analysis. Since \water{} and CO are the only detected molecules with high significance in our observations of CoRoT-2b, we constructed a planetary model featuring these two molecules. Furthermore, to elucidate the spatial distribution of \water{} and CO, we ran retrievals with atmospheric models including each molecule individually. We note that, generally, running single-species retrievals may not result in consistent abundances, because modelled line depths can be affected due to other missing opacity sources~\citep{brogi_equation_2019}.

\begin{figure}[!t]
    \centering
    \includegraphics[width=0.95\linewidth]{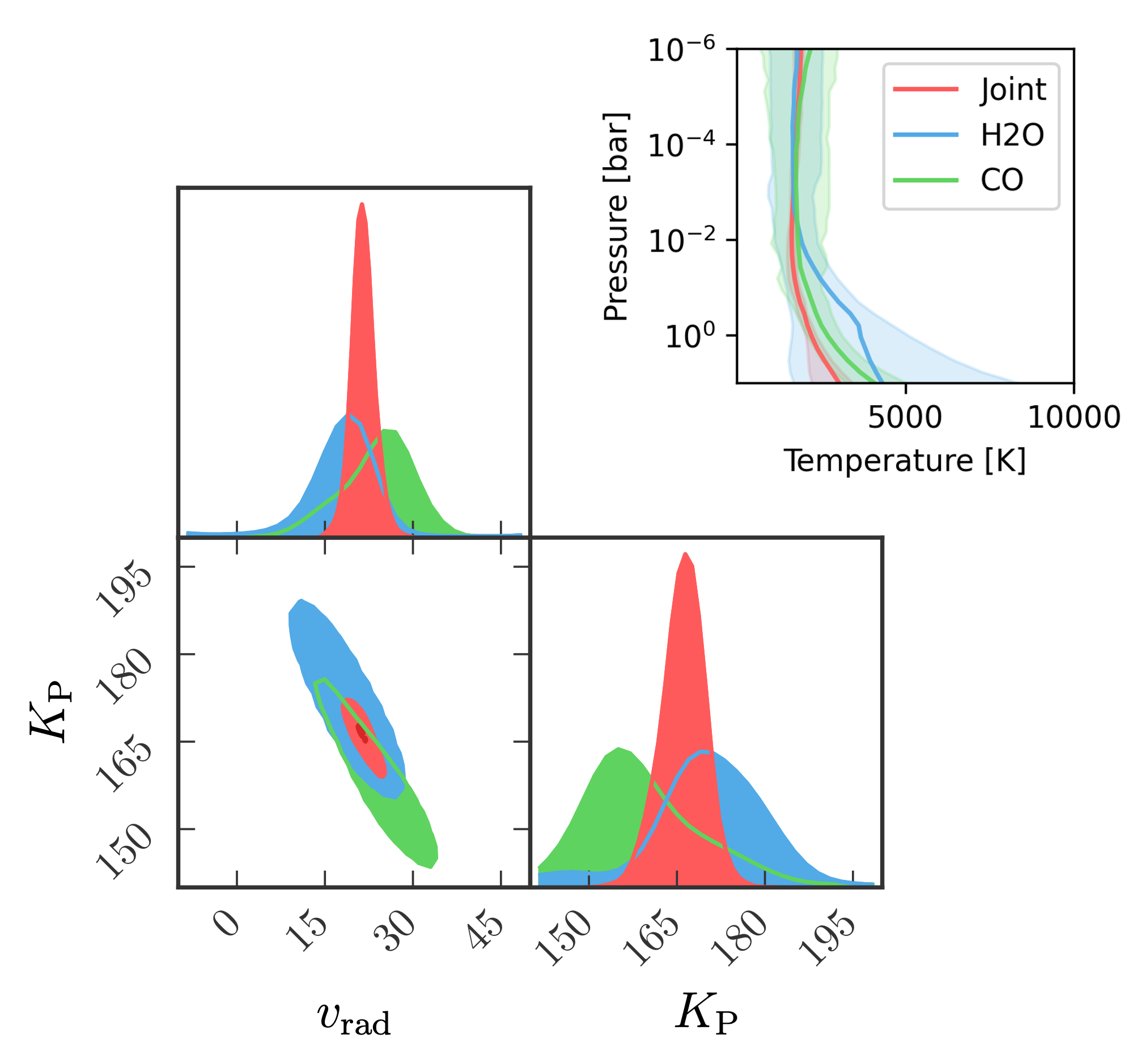}
    \caption{\emph{Top:} Overlay of retrieved CO and \water{} T-P profiles which shows that they are in agreement within 1$\sigma$.
    \emph{bottom:} The retrieved \kp{} and \vrad{} from three different retrievals, joint retrieval combined with low-resolution data, \water{}-only, and CO-only retrieval. The retrieved \kp{} and \vrad{} from all three retrievals are in agreement within 1$\sigma$.
    }
    \label{fig:small_corner}
\end{figure}

The retrieved results for the three sets of retrievals are depicted in Figure \ref{fig:small_corner}. The top figure presents the retrieved T-P profile of the full species retrieval and single-molecule retrieval. The bottom small corner plot illustrates the retrieved posterier distributions of \kp{} and \vrad{} for models with \water{} and CO separately, along with the joint model. The T-P profiles retrieved from three different retrievals are similar and are consistent within 1$\sigma$.
As expected, the molecular abundance of CO remains constant vertically as CO only dissociated at high temperature. As the three retrieved T-P profiles are all in agreement, this likely indicates that both CO and water probe similar pressure ranges. 

Any variations in \kp{} and \vrad{} could be attributed to planetary rotation or atmospheric winds, combined with heterogeneities in atmospheric properties, such as differing temperature-pressure profiles, abundances, or cloud cover across the visible hemisphere of CoRoT-2b.
When we ran a retrieval individually with \water{} and CO, we do not observe significant variations in \kp{} and $v_{\text{rad}}$. Specifically, for the model featuring \water{} exclusively, we obtained values of $171_{-26}  ^{+17}$ km s$^{-1}$ and $18.6_{-18.1}^{+12.1}$ km s$^{-1}$. Meanwhile, for the model containing CO alone, the values were $156_{-14}^{+24}$ km s$^{-1}$ and $25.2_{-14.7}^{+10.0}$ km s$^{-1}$. 
With all measurements agreeing within 1$\sigma$, our analysis suggests that at the level of precision of these observations, no significant difference in dynamics is measured. 

\section{Conclusions} \label{sec:conclusion}
In summary, we present the analysis of pre-eclipse observations using Gemini-S/IGRINS of the unusual hot Jupiter, CoRoT-2b and report the first high-resolution spectroscopic detection of molecular species in its atmosphere.  Overall our results provide new insights into the atmospheric composition and physical processes governing this peculiar hot Jupiter. Using two independent analyses, we revealed the presence of \water{} and CO in the pre-eclipse observations. Our observations also suggest a more complex atmospheric structure than previously inferred from previous analyses that described CoRoT-2b’s emission spectrum as being mostly featureless and well fit by a blackbody. Our initial cross-correlation analysis revealed tentative absorption signals for HCN and OH, however, our retrievals could not constrain their abundances.  We therefore cannot confirm or rule out the presence of HCN or OH in the dayside atmosphere of CoRoT-2b. No robust evidence was found for other molecules such as VO, TiO, CO$_2$, CH$_4$ and SiO.

Our observations also show a lack of prominent water features at wavelengths shorter than 1.7\,$\mu$m, observed in both IGRINS and WFC3 data. This could be due clouds on the dayside or the presence of H- potentially driven by enhanced UV radiation from the young and active host star which may also play a role in muting short-wavelength features. More pessimistically, the lack of features at short wavelength could be due to instrumental systematics affecting the WFC3 and IGRINS data. Additional spectroscopic observations will be required to determine the origin of this lack of molecular feature. Like WASP-77A b and HIP-65 A b, CoRoT-2b also shows evidence of having a sub-solar metallicity atmosphere and a solar to super-solar C/O ratio, suggesting that CoRoT-2b may have formed beyond the water ice line then migrating inward after the dissipation of the protoplanetary disk, without accreting a significant quantity of planetesimals. However, more observations are necessary to validate this scenario.

Additionally, single-molecule retrievals show no significant differences in the spatial distribution of \water{} and CO, suggesting a well-mixed atmosphere with minimal evidence for distinct atmospheric dynamics between the detected molecules.
This work represents the first part of a comprehensive investigation into the atmospheric properties and unusual dynamical features of CoRoT-2b, including its unique westward hotspot offset. Future studies will leverage phase-resolved observations with CRIRES+ to further probe the planet’s thermal structure and atmospheric circulation patterns. 

\newpage

\begin{acknowledgments}
Y.S. acknowledges support from the Institut Trottier de Recherche sur les Exoplanètes (iREx) for carrying out this work. L.D.\ is a Banting and Trottier Postdoctoral fellow and would like to acknowledge funding from the National Sciences and Research Council of Canada (NSERC), and the Institut Trottier de recherche sur les exoplanètes (iREx). L. D. would also like to support from L'Oréal Canada and the Canadian Commission for UNESCO's Research Excellence Fellowship. This project has been in part carried out within the framework of the National Centre of Competence in Research PlanetS supported by the Swiss National Science Foundation under grant 51NF40\_205606. The authors acknowledge the financial support of the SNSF. R.A. acknowledges the Swiss National Science Foundation (SNSF) support under the Post-Doc Mobility grant P500PT\_222212 and the support of the Institut Trottier de Recherche sur les Exoplanètes (iREx). A.S.L. acknowledges financial support from the Severo Ochoa grant CEX2021-001131-S funded by MCIN/AEI/ 10.13039/501100011033.
\end{acknowledgments}

\newpage
\clearpage

\appendix
\section{Water Dissociation Parameter in Planetary Model}
\label{sec:app_water_dissociation}

CoRoT-2b’s dayside atmosphere likely reaches temperatures high enough for key molecules such as \water{} to partially dissociate into their atomic constituents. To incorporate this behaviour in our atmospheric models, we followed the dissociation treatment introduced by \citep{parmentier_2018}. In this approach, the abundance profile $A$ is a combination of a constant abundance, $A_0$, and dissociated abundance, $A_d$, described by the expression:
\begin{equation}
    \frac{1}{A}=\left(\frac{1}{\sqrt{A_0}} + \frac{1}{\sqrt{A_d}}\right)^2 \,.
\end{equation}
The term $A_d$ follows a pressure-dependent power law, defined as
\begin{equation}
    A_d = 10^{-\gamma} P^\alpha 10^{\frac{\beta}{T}}\,,
\end{equation}
where $\gamma$, $\alpha$ and $\beta$ are free parameters specific to each species. To enable the profile to shift toward higher abundances, we added a dependence on $\mathrm{A}_0$. This modification affects only $\gamma$, whose updated form as a function of $\mathrm{A}_0$ is given by
\begin{equation}
    \gamma = \log_{10} \frac{\mathrm{A}_0}{\mathrm{A}_{\rm Ref}} - \gamma_{\rm Ref} \, .
\end{equation}

In the present analysis, this parametrization is applied only to water vapour, as \water{} provides the dominant molecular absorption across the IGRINS near-infrared wavelength range.

\newpage
\section{Atmospheric Retrieval Spectral Points and Example Corner
Plots}
\label{sec: corner plots}
During the joint retrieval, we combined the high-resolution
IGRINS spectra with previously published low-resolution
eclipse spectra obtained from HST and Spitzer. The low-resolution spectral data points are listed in Table~\ref{tab:low_res_points}.

\begin{figure*}[!t]
    \centering
    \includegraphics[width=\linewidth]{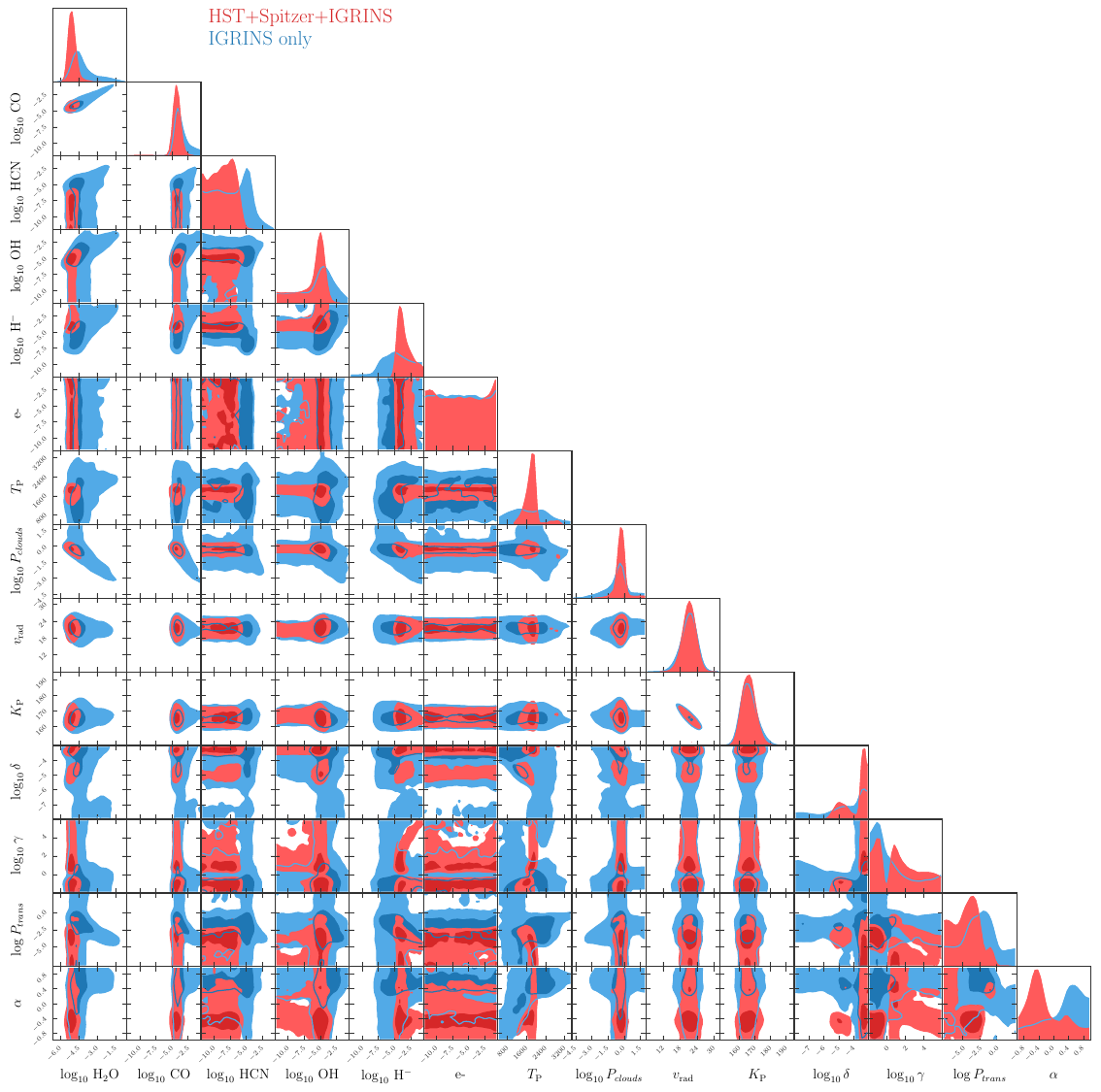}
    \caption{Corner plot for the free retrievals, showing only the molecules detected via cross correlation. The posterior distributions obtained from the retrieval using only the high-resolution IGRINS observations are shown in blue, while those from the retrieval combining the high-resolution data with low-resolution HST and Spitzer observations are shown in red.}
    \label{fig:corner_detected}
\end{figure*}

\begin{figure*}[!t]
    \centering
    \includegraphics[width=\linewidth]{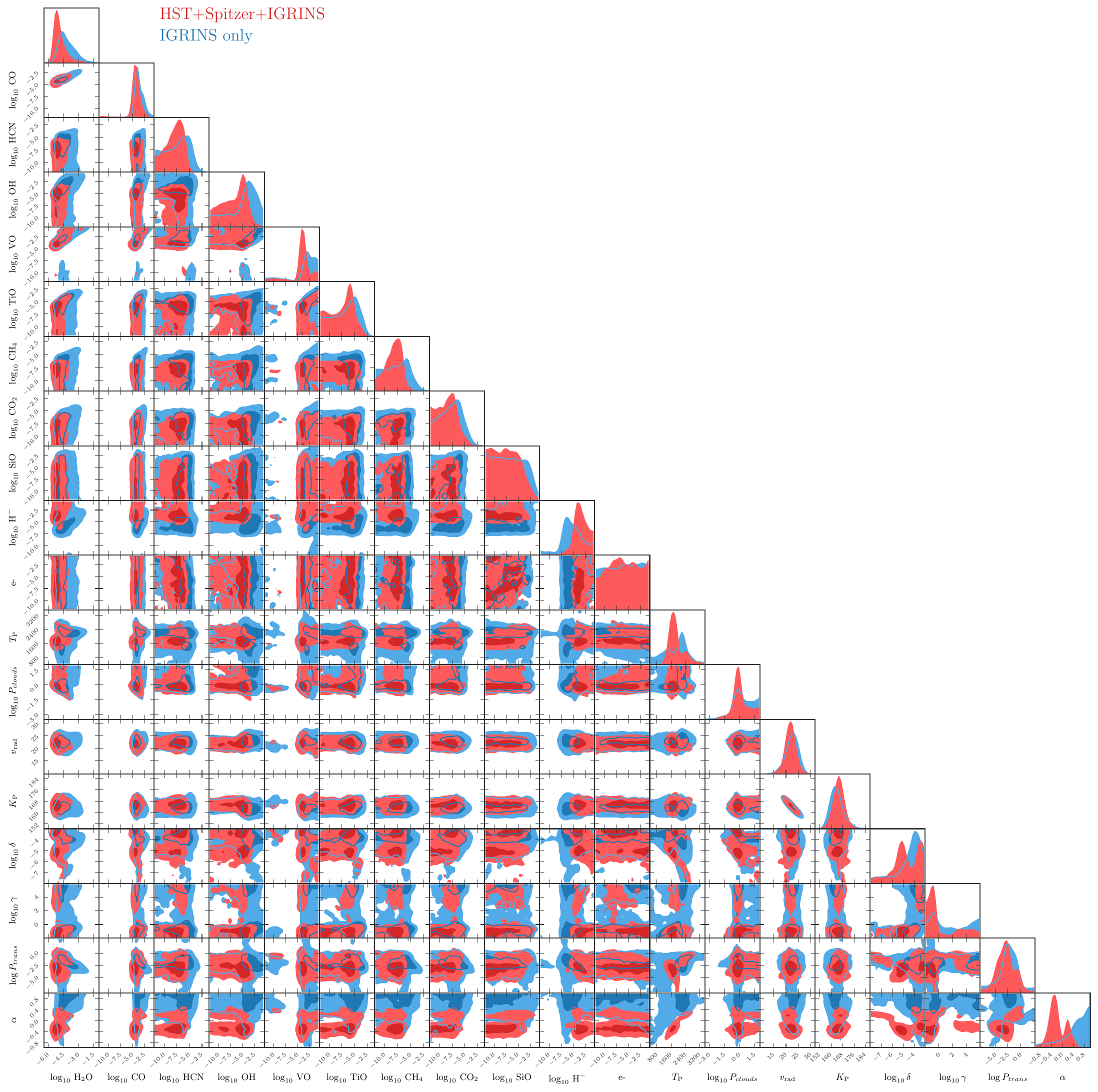}
    \caption{Corner plot for the free retrievals, showing both molecules detected via cross correlation and those that were not detected. The posterior distributions obtained from the retrieval using only the high-resolution IGRINS observations are shown in blue, while those from the retrieval combining the high-resolution data with low-resolution HST and Spitzer observations are shown in red.
    }
    \label{fig:corner_joint}
\end{figure*}

The resulting corner plots for the free retrievals, considering
(i) only molecules detected in the cross-correlation analysis and (ii) the full set of molecules, including both detected and non-detected species, are shown in Figures~\ref{fig:corner_detected} and \ref{fig:corner_joint}, respectively.

\begin{table}[ht]
    \centering
    \caption{Low-resolution eclipse spectral points used in the joint retrieval. The HST/WFC3 data for CoRoT-2b are from Wilkins et al.\ (2014), and the Spitzer data are from Deming et al.\ (2012).}
    \label{tab:low_res_points}
    \begin{tabular}{ccc}
        \hline
        \textbf{Wavelength ($\mu$m)} & \textbf{$F_p/F_\star$} & \textbf{Uncertainty} \\
        \hline
        \multicolumn{3}{c}{\textit{HST/WFC3} \citep{wilkins_pastobs_2014}} \\
        1.125 & 0.03346 & 0.00674 \\
        1.169 & 0.02724 & 0.00837 \\
        1.218 & 0.03394 & 0.01193 \\
        1.278 & 0.03442 & 0.00720 \\
        1.324 & 0.03389 & 0.00647 \\
        1.369 & 0.04039 & 0.00771 \\
        1.424 & 0.04545 & 0.00598 \\
        1.475 & 0.03203 & 0.00935 \\
        1.525 & 0.04383 & 0.00626 \\
        1.574 & 0.05487 & 0.00613 \\
        1.619 & 0.03822 & 0.00822 \\
        \hline
        \multicolumn{3}{c}{\textit{Spitzer} \citep{Deming_pastobs_2011}} \\
        4.5 & 0.00488 & 0.00020 \\
        \hline
    \end{tabular}
\end{table}

\newpage
\clearpage
\section{Independent Analysis of IGRINS} \label{sec:app_indep}

To further confirm the results, we perform an independent analysis of the high-resolution IGRINS observations of CoRoT-2b.  For this, we use the same outputted data from the IGRINS pipeline, however all subsequent modeling and detrending are completely different compared to the main retrievals presented in the main text. 

For the data processing, we use a PCA-based procedure following the steps explained in \citet{bazinet_highres_2024}, to which the reader is referred to for more detail~\citep[see also][]{pelletier_where_2021, Pelletier_2023, pelletier_crires_2024}. 
For the modeling, we use the SCARLET framework \citep{benneke_atmospheric_2012, benneke_how_2013, benneke_strict_2015, pelletier_where_2021} to generate synthetic planetary spectra.

Similar to the main analysis, we run a free retrieval fitting the log volume mixing ratios of \water, CO, HCN, OH, CH$_4$ and CO$_2$, with all abundances assumed to be constant with altitude. We also fit the orbital parameters \kp{} and $v_{\text{rad}}$ as well as a cloud top pressure.
However, we fit the vertical temperature profile freely as in \cite{pelletier_where_2021}, using 15 equally distant points in log-pressure space between $10^2$ and $10^{-5}$ bars with a smoothing prior placed on the temperature profile to penalize unphysical jagged profiles.

\begin{figure*}[!htb]
    \centering
    \includegraphics[width=\linewidth]{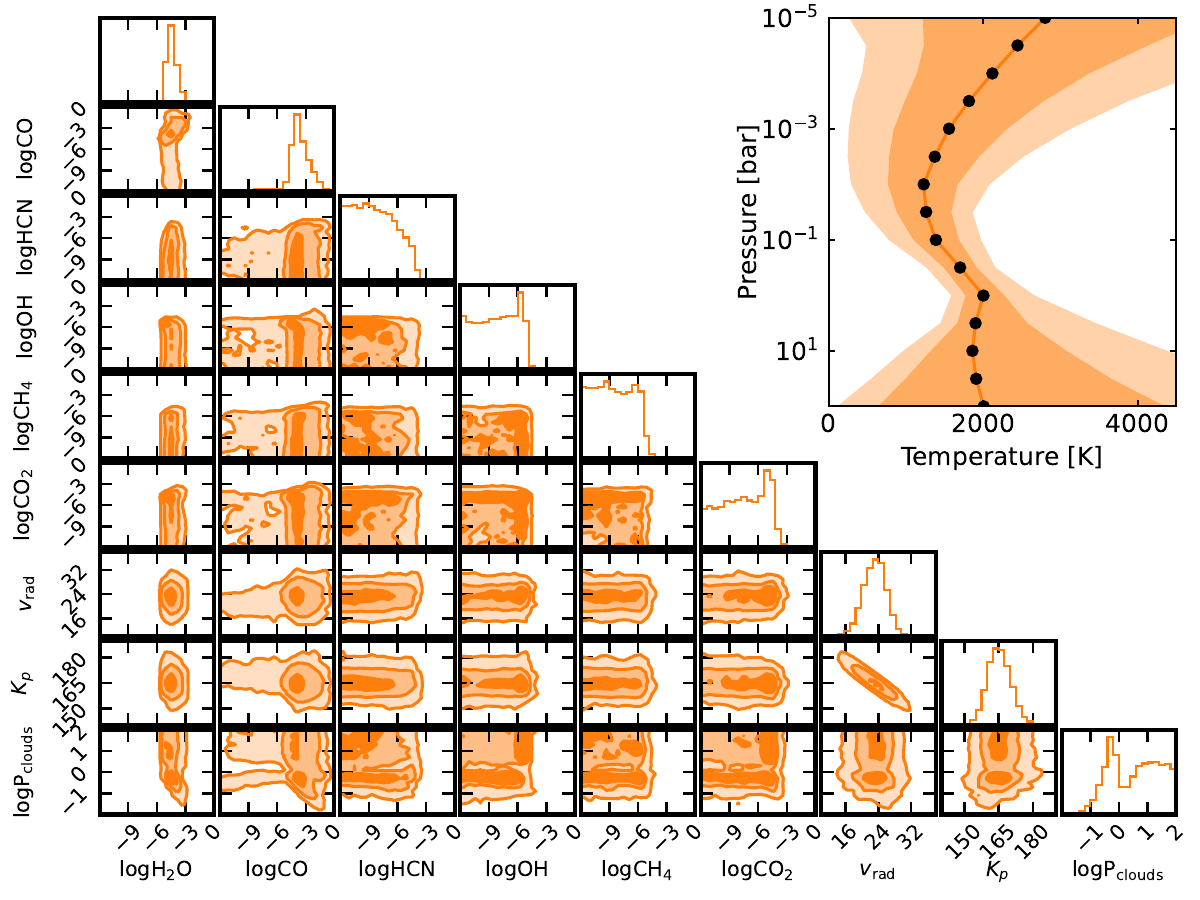}
    \caption{Corner plot and vertical temperature profile for the atmospheric parameters of CoRoT-2b determined from the independent analysis (Appendix~\ref{sec:app_indep}).  Of the molecules, only bounded constraints are obtained for H$_2$O, with the CO posterior still being consistent with being null at 3$\sigma$ and only upper limits obtained for HCN, OH, CO$_2$, and CH$_4$.  The retrieved temperature structure is non-inverted in the regions probed by the data (where the uncertainties of the TP profile are narrowest).  The apparent rise in temperature at high altitudes is likely not real and rather a result of the lack of sensitivity in those regions combined with the temperature prior allowing only positive values.  }
    \label{fig:retrieval_scarlet}
\end{figure*}

\begin{table}[!htb]
    \centering
    \caption{Results from the SCARLET retrieval compared to the retrieval presented in the main article. The upper and lower limits are 2-$\sigma$ limits.}
    \begin{tabular}{c c c c}
         Parameter & SCARLET prior & SCARLET retrieval & Full species retrieval  \\
         \hline
         log$_{10}$H$_2$O & U(-12, 0) & $-4.49^{+0.61}_{-0.43}$ & $-4.50^{+0.72}_{-0.60}$ \\ 
         log$_{10}$CO & U(-12, 0) & $-3.77^{+1.16}_{-0.79}$ & $-3.79^{+0.79}_{-0.69}$ \\ 
         log$_{10}$HCN & U(-12, 0) & $< -4.76$ & $< -4.11$ \\
         log$_{10}$OH & U(-12, 0) & $< -5.25$ & $< -0.50$ \\ 
         log$_{10}$CH$_4$ & U(-12, 0) & $< -5.55$ & $< -4.34$ \\ 
         log$_{10}$CO$_2$ & U(-12, 0) & $< -4.36$ & $< -4.46$ \\ 
         log$_{10}$P$_\mathrm{clouds}$ & U(-6, 2) & $> -0.75$ & $-0.06^{+1.39}_{-0.63}$ \\
         \vrad & U(5, 45) & $23.19^{+2.58}_{-2.92}$ & $22.07^{+2.43}_{-2.52}$ \\ 
         \kp & U(144.7, 204.7) & $165.17^{+5.26}_{-4.83}$ & $164.18^{+4.34}_{-4.38}$ \\ 
         \hline
    \end{tabular}
    \label{tab:scarlet_vs_main}
\end{table}

From this analysis, we find overall consistent results as presented in the main text. In particular, \water\ and CO are retrieved with almost identical median mixing ratios between both analyses (Table~\ref{tab:scarlet_vs_main}). However, the CO posterior does show a tail extending to low abundances, reflecting the weaker cross-correlation detection relative to H$_2$O.  Models with HCN, OH, CO$_2$, or CH$_4$ are not preferred, with only upper limits being obtained.

\newpage
\clearpage
\bibliography{sample631}{}

@ARTICLE{line_wasp77_2021,
       author = {{Line}, Michael R. and {Brogi}, Matteo and {Bean}, Jacob L. and {Gandhi}, Siddharth and {Zalesky}, Joseph and {Parmentier}, Vivien and {Smith}, Peter and {Mace}, Gregory N. and {Mansfield}, Megan and {Kempton}, Eliza M. -R. and {Fortney}, Jonathan J. and {Shkolnik}, Evgenya and {Patience}, Jennifer and {Rauscher}, Emily and {D{\'e}sert}, Jean-Michel and {Wardenier}, Joost P.},
        title = "{A solar C/O and sub-solar metallicity in a hot Jupiter atmosphere}",
      journal = {\nat},
         year = 2021,
        month = oct,
       volume = {598},
       number = {7882},
        pages = {580-584},
          doi = {10.1038/s41586-021-03912-6},
archivePrefix = {arXiv},
       eprint = {2110.14821},
 primaryClass = {astro-ph.EP},
       adsurl = {https://ui.adsabs.harvard.edu/abs/2021Natur.598..580L}
}

@ARTICLE{boucher_co_2023,
       author = {{Boucher}, Anne and {Lafreni{\'e}re}, David and {Pelletier}, Stefan and {Darveau-Bernier}, Antoine and {Radica}, Michael and {Allart}, Romain and {Artigau}, {\'E}tienne and {Cook}, Neil J. and {Debras}, Florian and {Doyon}, Ren{\'e} and {Gaidos}, Eric and {Benneke}, Bj{\"o}rn and {Cadieux}, Charles and {Carmona}, Andres and {Cloutier}, Ryan and {Cort{\'e}s-Zuleta}, P{\'\i}a and {Cowan}, Nicolas B. and {Delfosse}, Xavier and {Donati}, Jean-Fran{\c{c}}ois and {Fouqu{\'e}}, Pascal and {Forveille}, Thierry and {Grankin}, Konstantin and {H{\'e}brard}, Guillaume and {Martins}, Jorge H.~C. and {Martioli}, Eder and {Masson}, Adrien and {Vinatier}, Sandrine},
        title = "{CO or no CO? Narrowing the CO abundance constraint and recovering the H$_{2}$O detection in the atmosphere of WASP-127 b using SPIRou}",
      journal = {\mnras},
         year = 2023,
        month = jul,
       volume = {522},
       number = {4},
        pages = {5062-5083},
          doi = {10.1093/mnras/stad1247},
archivePrefix = {arXiv},
       eprint = {2303.03232},
 primaryClass = {astro-ph.EP},
       adsurl = {https://ui.adsabs.harvard.edu/abs/2023MNRAS.522.5062B}
}

@ARTICLE{boucher_reduction_2021,
       author = {{Boucher}, Anne and {Darveau-Bernier}, Antoine and {Pelletier}, Stefan and {Lafreni{\`e}re}, David and {Artigau}, {\'E}tienne and {Cook}, Neil J. and {Allart}, Romain and {Radica}, Michael and {Doyon}, Ren{\'e} and {Benneke}, Bj{\"o}rn and {Arnold}, Luc and {Bonfils}, Xavier and {Bourrier}, Vincent and {Cloutier}, Ryan and {Gomes da Silva}, Jo{\~a}o and {Deibert}, Emily and {Delfosse}, Xavier and {Donati}, Jean-Fran{\c{c}}ois and {Ehrenreich}, David and {Figueira}, Pedro and {Forveille}, Thierry and {Fouqu{\'e}}, Pascal and {Gagn{\'e}}, Jonathan and {Gaidos}, Eric and {H{\'e}brard}, Guillaume and {Jayawardhana}, Ray and {Klein}, Baptiste and {Lovis}, Christophe and {Martins}, Jorge H.~C. and {Martioli}, Eder and {Moutou}, Claire and {Santos}, Nuno C.},
        title = "{Characterizing Exoplanetary Atmospheres at High Resolution with SPIRou: Detection of Water on HD 189733 b}",
      journal = {\aj},
         year = 2021,
        month = dec,
       volume = {162},
       number = {6},
          eid = {233},
        pages = {233},
          doi = {10.3847/1538-3881/ac1f8e},
archivePrefix = {arXiv},
       eprint = {2108.08390},
 primaryClass = {astro-ph.EP},
       adsurl = {https://ui.adsabs.harvard.edu/abs/2021AJ....162..233B}
}

@ARTICLE{Dang_c2b_2018,
       author = {{Dang}, Lisa and {Cowan}, Nicolas B. and {Schwartz}, Joel C. and {Rauscher}, Emily and {Zhang}, Michael and {Knutson}, Heather A. and {Line}, Michael and {Dobbs-Dixon}, Ian and {Deming}, Drake and {Sundararajan}, Sudarsan and {Fortney}, Jonathan J. and {Zhao}, Ming},
        title = "{Detection of a westward hotspot offset in the atmosphere of hot gas giant CoRoT-2b}",
      journal = {Nature Astronomy},
         year = 2018,
        month = mar,
       volume = {2},
        pages = {220-227},
          doi = {10.1038/s41550-017-0351-6},
archivePrefix = {arXiv},
       eprint = {1801.06548},
 primaryClass = {astro-ph.EP},
       adsurl = {https://ui.adsabs.harvard.edu/abs/2018NatAs...2..220D}
}

@ARTICLE{madhusudhan_overview_2019,
       author = {{Madhusudhan}, Nikku},
        title = "{Exoplanetary Atmospheres: Key Insights, Challenges, and Prospects}",
      journal = {\araa},
         year = 2019,
        month = aug,
       volume = {57},
        pages = {617-663},
          doi = {10.1146/annurev-astro-081817-051846},
archivePrefix = {arXiv},
       eprint = {1904.03190},
 primaryClass = {astro-ph.EP},
       adsurl = {https://ui.adsabs.harvard.edu/abs/2019ARA&A..57..617M}
}

@incollection{brogi_highres_2021,
	author = {Brogi, Matteo and Birkby, Jayne},
	booktitle = {ExoFrontiers},
	doi = {10.1088/2514-3433/abfa8fch8},
	isbn = {978-0-7503-1472-5},
	pages = {8-1 to 8-10},
	publisher = {IOP Publishing},
	series = {2514-3433},
	title = {High-resolution Spectroscopy},
	type = {Book Chapter},
	url = {https://dx.doi.org/10.1088/2514-3433/abfa8fch8},
	year = {2021}}

@ARTICLE{bazinet_highres_2024,
       author = {{Bazinet}, Luc and {Pelletier}, Stefan and {Benneke}, Bj{\"o}rn and {Salinas}, Ricardo and {Mace}, Gregory N.},
        title = "{A sub-solar metallicity on the ultra-short period planet HIP 65Ab}",
      journal = {arXiv e-prints},
         year = 2024,
        month = mar,
          eid = {arXiv:2403.07983},
        pages = {arXiv:2403.07983},
archivePrefix = {arXiv},
       eprint = {2403.07983},
 primaryClass = {astro-ph.EP},
       adsurl = {https://ui.adsabs.harvard.edu/abs/2024arXiv240307983B}
}

@ARTICLE{seager_atmosphere_2010,
       author = {{Seager}, Sara and {Deming}, Drake},
        title = "{Exoplanet Atmospheres}",
      journal = {\araa},
         year = 2010,
        month = sep,
       volume = {48},
        pages = {631-672},
          doi = {10.1146/annurev-astro-081309-130837},
archivePrefix = {arXiv},
       eprint = {1005.4037},
 primaryClass = {astro-ph.EP},
       adsurl = {https://ui.adsabs.harvard.edu/abs/2010ARA&A..48..631S}
}

@ARTICLE{fortney_hotJup_2021,
       author = {{Fortney}, Jonathan J. and {Dawson}, Rebekah I. and {Komacek}, Thaddeus D.},
        title = "{Hot Jupiters: Origins, Structure, Atmospheres}",
      journal = {Journal of Geophysical Research (Planets)},
         year = 2021,
        month = mar,
       volume = {126},
       number = {3},
          eid = {e06629},
        pages = {e06629},
          doi = {10.1029/2020JE006629},
archivePrefix = {arXiv},
       eprint = {2102.05064},
 primaryClass = {astro-ph.EP},
       adsurl = {https://ui.adsabs.harvard.edu/abs/2021JGRE..12606629F}
}

@ARTICLE{encrenaz_infrared_2014,
       author = {{Encrenaz}, T.},
        title = "{Infrared spectroscopy of exoplanets: observational constraints}",
      journal = {Philosophical Transactions of the Royal Society of London Series A},
         year = 2014,
        month = mar,
       volume = {372},
       number = {2014},
        pages = {20130083-20130083},
          doi = {10.1098/rsta.2013.0083},
       adsurl = {https://ui.adsabs.harvard.edu/abs/2014RSPTA.37230083E}
}

@ARTICLE{may_exoplanet_2022,
       author = {{May}, E.~M. and {Stevenson}, K.~B. and {Bean}, Jacob L. and {Bell}, Taylor J. and {Cowan}, Nicolas B. and {Dang}, Lisa and {Desert}, Jean-Michel and {Fortney}, Jonathan J. and {Keating}, Dylan and {Kempton}, Eliza M. -R. and {Komacek}, Thaddeus D. and {Lewis}, Nikole K. and {Mansfield}, Megan and {Morley}, Caroline and {Parmentier}, Vivien and {Rauscher}, Emily and {Swain}, Mark R. and {Zellem}, Robert T. and {Showman}, Adam},
        title = "{A New Analysis of Eight Spitzer Phase Curves and Hot Jupiter Population Trends: Qatar-1b, Qatar-2b, WASP-52b, WASP-34b, and WASP-140b}",
      journal = {\aj},
         year = 2022,
        month = jun,
       volume = {163},
       number = {6},
          eid = {256},
        pages = {256},
          doi = {10.3847/1538-3881/ac6261},
archivePrefix = {arXiv},
       eprint = {2203.15059},
 primaryClass = {astro-ph.EP},
       adsurl = {https://ui.adsabs.harvard.edu/abs/2022AJ....163..256M}
}

@ARTICLE{parmentier_2018,
       author = {{Parmentier}, Vivien and {Line}, Mike R. and {Bean}, Jacob L. and {Mansfield}, Megan and {Kreidberg}, Laura and {Lupu}, Roxana and {Visscher}, Channon and {D{\'e}sert}, Jean-Michel and {Fortney}, Jonathan J. and {Deleuil}, Magalie and {Arcangeli}, Jacob and {Showman}, Adam P. and {Marley}, Mark S.},
        title = "{From thermal dissociation to condensation in the atmospheres of ultra hot Jupiters: WASP-121b in context}",
      journal = {\aap},
         year = 2018,
        month = sep,
       volume = {617},
          eid = {A110},
        pages = {A110},
          doi = {10.1051/0004-6361/201833059},
archivePrefix = {arXiv},
       eprint = {1805.00096},
 primaryClass = {astro-ph.EP},
       adsurl = {https://ui.adsabs.harvard.edu/abs/2018A&A...617A.110P}
}

@ARTICLE{bell_spitzer_2021,
       author = {{Bell}, Taylor J. and {Dang}, Lisa and {Cowan}, Nicolas B. and {Bean}, Jacob and {D{\'e}sert}, Jean-Michel and {Fortney}, Jonathan J. and {Keating}, Dylan and {Kempton}, Eliza and {Kreidberg}, Laura and {Line}, Michael R. and {Mansfield}, Megan and {Parmentier}, Vivien and {Stevenson}, Kevin B. and {Swain}, Mark and {Zellem}, Robert T.},
        title = "{A comprehensive reanalysis of Spitzer's 4.5 {\ensuremath{\mu}}m phase curves, and the phase variations of the ultra-hot Jupiters MASCARA-1b and KELT-16b}",
      journal = {\mnras},
         year = 2021,
        month = jul,
       volume = {504},
       number = {3},
        pages = {3316-3337},
          doi = {10.1093/mnras/stab1027},
archivePrefix = {arXiv},
       eprint = {2010.00687},
 primaryClass = {astro-ph.EP},
       adsurl = {https://ui.adsabs.harvard.edu/abs/2021MNRAS.504.3316B}
}

@ARTICLE{mraz_2024,
       author = {{Mraz}, Georgia and {Darveau-Bernier}, Antoine and {Boucher}, Anne and {Cowan}, Nicolas B. and {Lafreni{\`e}re}, David and {Cadieux}, Charles},
        title = "{Out of the Darkness: High-resolution Detection of CO Absorption on the Nightside of WASP-33b}",
      journal = {\apjl},
         year = 2024,
        month = nov,
       volume = {975},
       number = {2},
          eid = {L42},
        pages = {L42},
          doi = {10.3847/2041-8213/ad8438},
archivePrefix = {arXiv},
       eprint = {2410.11060},
 primaryClass = {astro-ph.EP},
       adsurl = {https://ui.adsabs.harvard.edu/abs/2024ApJ...975L..42M}
}

@article{guillot_c_age_2011,
	author = {{Guillot, T.} and {Havel, M.}},
	doi = {10.1051/0004-6361/201015051},
	journal = {A\&A},
	pages = {A20},
	title = {An analysis of the CoRoT-2 system: a young spotted star and its inflated giant planet},
	url = {https://doi.org/10.1051/0004-6361/201015051},
	volume = 527,
	year = 2011
}

@article{pelletier_where_2021,
	title = {Where {Is} the {Water}? {Jupiter}-like {C}/{H} {Ratio} but {Strong} {H2O} {Depletion} {Found} on τ {Boötis} b {Using} {SPIRou}},
	volume = {162},
	issn = {1538-3881},
	shorttitle = {Where {Is} the {Water}?},
	doi = {10.3847/1538-3881/ac0428},
	language = {en},
	number = {2},
	urldate = {2021-08-17},
	journal = {The Astronomical Journal},
	author = {Pelletier, Stefan and Benneke, Björn and Darveau-Bernier, Antoine and Boucher, Anne and Cook, Neil J. and Piaulet, Caroline and Coulombe, Louis-Philippe and Artigau, Étienne and Lafrenière, David and Delisle, Simon and Allart, Romain and Doyon, René and Donati, Jean-François and Fouqué, Pascal and Moutou, Claire and Cadieux, Charles and Delfosse, Xavier and Hébrard, Guillaume and Martins, Jorge H. C. and Martioli, Eder and Vandal, Thomas},
	month = jul,
	year = {2021},
	pages = {73},
}

@article{yan_crires_2023,
	title = {{CRIRES}+ detection of {CO} emissions lines and temperature inversions on the dayside of {WASP}-18b and {WASP}-76b},
	volume = {672},
	copyright = {© The Authors 2023},
	issn = {0004-6361, 1432-0746},
	url = {https://www.aanda.org/articles/aa/abs/2023/04/aa45371-22/aa45371-22.html},
	doi = {10.1051/0004-6361/202245371},
	language = {en},
	urldate = {2023-08-28},
	journal = {Astronomy \& Astrophysics},
	author = {Yan, F. and Nortmann, L. and Reiners, A. and Piskunov, N. and Hatzes, A. and Seemann, U. and Shulyak, D. and Lavail, A. and Rains, A. D. and Cont, D. and Rengel, M. and Lesjak, F. and Nagel, E. and Kochukhov, O. and Czesla, S. and Boldt-Christmas, L. and Heiter, U. and Smoker, J. V. and Rodler, F. and Bristow, P. and Dorn, R. J. and Jung, Y. and Marquart, T. and Stempels, E.},
	month = apr,
	year = {2023},
	pages = {A107}
    }

@article{Baluev_2015,
       author = {{Baluev}, Roman V. and {Sokov}, Evgenii N. and {Shaidulin}, Vakhit Sh. and {Sokova}, Iraida A. and {Jones}, Hugh R.~A. and {Tuomi}, Mikko and {Anglada-Escud{\'e}}, Guillem and {Benni}, Paul and {Colazo}, Carlos A. and {Schneiter}, Matias E. and {Villarreal D'Angelo}, Carolina S. and {Burdanov}, Artem Yu. and {Fern{\'a}ndez-Laj{\'u}s}, Eduardo and {Ba{\c{s}}t{\"u}rk}, {\"O}zg{\"u}r and {Hentunen}, Veli-Pekka and {Shadick}, Stan},
        title = "{Benchmarking the power of amateur observatories for TTV exoplanets detection}",
      journal = {\mnras},
         year = 2015,
        month = jul,
       volume = {450},
       number = {3},
        pages = {3101-3113},
          doi = {10.1093/mnras/stv788},
archivePrefix = {arXiv},
       eprint = {1501.06748},
 primaryClass = {astro-ph.IM},
       adsurl = {https://ui.adsabs.harvard.edu/abs/2015MNRAS.450.3101B}
}

@ARTICLE{kokori_c2b_2022,
       author = {{Kokori}, A. and {Tsiaras}, A. and {Edwards}, B. and {Rocchetto}, M. and {Tinetti}, G. and {Bewersdorff}, L. and {Jongen}, Y. and {Lekkas}, G. and {Pantelidou}, G. and {Poultourtzidis}, E. and {W{\"u}nsche}, A. and {Aggelis}, C. and {Agnihotri}, V.~K. and {Arena}, C. and {Bachschmidt}, M. and {Bennett}, D. and {Benni}, P. and {Bernacki}, K. and {Besson}, E. and {Betti}, L. and {Biagini}, A. and {Brandebourg}, P. and {Bretton}, M. and {Brincat}, S.~M. and {Cal{\'o}}, M. and {Campos}, F. and {Casali}, R. and {Ciantini}, R. and {Crow}, M.~V. and {Dauchet}, B. and {Dawes}, S. and {Deldem}, M. and {Deligeorgopoulos}, D. and {Dymock}, R. and {Eenm{\"a}e}, T. and {Evans}, P. and {Esseiva}, N. and {Falco}, C. and {Ferratfiat}, S. and {Fowler}, M. and {Futcher}, S.~R. and {Gaitan}, J. and {Horta}, F. Grau and {Guerra}, P. and {Hurter}, F. and {Jones}, A. and {Kang}, W. and {Kiiskinen}, H. and {Kim}, T. and {Laloum}, D. and {Lee}, R. and {Lomoz}, F. and {Lopresti}, C. and {Mallonn}, M. and {Mannucci}, M. and {Marino}, A. and {Mario}, J. -C. and {Marquette}, J. -B. and {Michelet}, J. and {Miller}, M. and {Mollier}, T. and {Molina}, D. and {Montigiani}, N. and {Mortari}, F. and {Morvan}, M. and {Mugnai}, L.~V. and {Naponiello}, L. and {Nastasi}, A. and {Neito}, R. and {Pace}, E. and {Papadeas}, P. and {Paschalis}, N. and {Pereira}, C. and {Perroud}, V. and {Phillips}, M. and {Pintr}, P. and {Pioppa}, J. -B. and {Popowicz}, A. and {Raetz}, M. and {Regembal}, F. and {Rickard}, K. and {Roberts}, M. and {Rousselot}, L. and {Rubia}, X. and {Savage}, J. and {Sedita}, D. and {Shave-Wall}, D. and {Sioulas}, N. and {{\v{S}}koln{\'\i}k}, V. and {Smith}, M. and {St-Gelais}, D. and {Stouraitis}, D. and {Strikis}, I. and {Thurston}, G. and {Tomacelli}, A. and {Tomatis}, A. and {Trevan}, B. and {Valeau}, P. and {Vignes}, J. -P. and {Vora}, K. and {Vra{\v{s}}{\v{t}}{\'a}k}, M. and {Walter}, F. and {Wenzel}, B. and {Wright}, D.~E. and {Z{\'\i}bar}, M.},
        title = "{ExoClock Project. II. A Large-scale Integrated Study with 180 Updated Exoplanet Ephemerides}",
      journal = {\apjs},
         year = 2022,
        month = feb,
       volume = {258},
       number = {2},
          eid = {40},
        pages = {40},
          doi = {10.3847/1538-4365/ac3a10},
archivePrefix = {arXiv},
       eprint = {2110.13863},
 primaryClass = {astro-ph.EP},
       adsurl = {https://ui.adsabs.harvard.edu/abs/2022ApJS..258...40K}
}

@ARTICLE{ozturk_c2b_2019,
       author = {{{\"O}zt{\"u}rk}, O{\v{g}}uz and {Erdem}, Ahmet},
        title = "{New photometric analysis of five exoplanets: CoRoT-2b, HAT-P-12b, TrES-2b, WASP-12b, and WASP-52b}",
      journal = {\mnras},
         year = 2019,
        month = jun,
       volume = {486},
       number = {2},
        pages = {2290-2307},
          doi = {10.1093/mnras/stz747},
       adsurl = {https://ui.adsabs.harvard.edu/abs/2019MNRAS.486.2290O}
}

@ARTICLE{guillon_pparameter_2010,
       author = {{Gillon}, M. and {Lanotte}, A.~A. and {Barman}, T. and {Miller}, N. and {Demory}, B. -O. and {Deleuil}, M. and {Montalb{\'a}n}, J. and {Bouchy}, F. and {Collier Cameron}, A. and {Deeg}, H.~J. and {Fortney}, J.~J. and {Fridlund}, M. and {Harrington}, J. and {Magain}, P. and {Moutou}, C. and {Queloz}, D. and {Rauer}, H. and {Rouan}, D. and {Schneider}, J.},
        title = "{The thermal emission of the young and massive planet CoRoT-2b at 4.5 and 8 {\ensuremath{\mu}}m}",
      journal = {\aap},
         year = 2010,
        month = feb,
       volume = {511},
          eid = {A3},
        pages = {A3},
          doi = {10.1051/0004-6361/200913507},
archivePrefix = {arXiv},
       eprint = {0911.5087},
 primaryClass = {astro-ph.EP},
       adsurl = {https://ui.adsabs.harvard.edu/abs/2010A&A...511A...3G}
}

@ARTICLE{wilkins_pastobs_2014,
       author = {{Wilkins}, Ashlee N. and {Deming}, Drake and {Madhusudhan}, Nikku and {Burrows}, Adam and {Knutson}, Heather and {McCullough}, Peter and {Ranjan}, Sukrit},
        title = "{The Emergent 1.1-1.7 {\ensuremath{\mu}}m Spectrum of the Exoplanet CoRoT-2b as Measured Using the Hubble Space Telescope}",
      journal = {\apj},
         year = 2014,
        month = mar,
       volume = {783},
       number = {2},
          eid = {113},
        pages = {113},
          doi = {10.1088/0004-637X/783/2/113},
archivePrefix = {arXiv},
       eprint = {1401.4464},
 primaryClass = {astro-ph.EP},
       adsurl = {https://ui.adsabs.harvard.edu/abs/2014ApJ...783..113W}
}

@ARTICLE{alonso_pastobs_2008,
       author = {{Alonso}, R. and {Auvergne}, M. and {Baglin}, A. and {Ollivier}, M. and {Moutou}, C. and {Rouan}, D. and {Deeg}, H.~J. and {Aigrain}, S. and {Almenara}, J.~M. and {Barbieri}, M. and {Barge}, P. and {Benz}, W. and {Bord{\'e}}, P. and {Bouchy}, F. and {de La Reza}, R. and {Deleuil}, M. and {Dvorak}, R. and {Erikson}, A. and {Fridlund}, M. and {Gillon}, M. and {Gondoin}, P. and {Guillot}, T. and {Hatzes}, A. and {H{\'e}brard}, G. and {Kabath}, P. and {Jorda}, L. and {Lammer}, H. and {L{\'e}ger}, A. and {Llebaria}, A. and {Loeillet}, B. and {Magain}, P. and {Mayor}, M. and {Mazeh}, T. and {P{\"a}tzold}, M. and {Pepe}, F. and {Pont}, F. and {Queloz}, D. and {Rauer}, H. and {Shporer}, A. and {Schneider}, J. and {Stecklum}, B. and {Udry}, S. and {Wuchterl}, G.},
        title = "{Transiting exoplanets from the CoRoT space mission. II. CoRoT-Exo-2b: a transiting planet around an active G star}",
      journal = {\aap},
         year = 2008,
        month = may,
       volume = {482},
       number = {3},
        pages = {L21-L24},
          doi = {10.1051/0004-6361:200809431},
archivePrefix = {arXiv},
       eprint = {0803.3207},
 primaryClass = {astro-ph},
       adsurl = {https://ui.adsabs.harvard.edu/abs/2008A&A...482L..21A}
}

@INPROCEEDINGS{baglin_c2b_2006,
       author = {{Baglin}, A. and {Auvergne}, M. and {Boisnard}, L. and {Lam-Trong}, T. and {Barge}, P. and {Catala}, C. and {Deleuil}, M. and {Michel}, E. and {Weiss}, W.},
        title = "{CoRoT: a high precision photometer for stellar ecolution and exoplanet finding}",
    booktitle = {36th COSPAR Scientific Assembly},
         year = 2006,
       volume = {36},
        month = jan,
        pages = {3749},
       adsurl = {https://ui.adsabs.harvard.edu/abs/2006cosp...36.3749B}
}

@ARTICLE{alonso_pastobs_2010,
       author = {{Alonso}, R. and {Deeg}, H.~J. and {Kabath}, P. and {Rabus}, M.},
        title = "{Ground-based Near-infrared Observations of the Secondary Eclipse of CoRoT-2b}",
      journal = {\aj},
         year = 2010,
        month = apr,
       volume = {139},
       number = {4},
        pages = {1481-1485},
          doi = {10.1088/0004-6256/139/4/1481},
archivePrefix = {arXiv},
       eprint = {1001.3060},
 primaryClass = {astro-ph.EP},
       adsurl = {https://ui.adsabs.harvard.edu/abs/2010AJ....139.1481A}
}

@ARTICLE{Deming_pastobs_2011,
       author = {{Deming}, Drake and {Knutson}, Heather and {Agol}, Eric and {Desert}, Jean-Michel and {Burrows}, Adam and {Fortney}, Jonathan J. and {Charbonneau}, David and {Cowan}, Nicolas B. and {Laughlin}, Gregory and {Langton}, Jonathan and {Showman}, Adam P. and {Lewis}, Nikole K.},
        title = "{Warm Spitzer Photometry of the Transiting Exoplanets CoRoT-1 and CoRoT-2 at Secondary Eclipse}",
      journal = {\apj},
         year = 2011,
        month = jan,
       volume = {726},
       number = {2},
          eid = {95},
        pages = {95},
          doi = {10.1088/0004-637X/726/2/95},
archivePrefix = {arXiv},
       eprint = {1011.1019},
 primaryClass = {astro-ph.EP},
       adsurl = {https://ui.adsabs.harvard.edu/abs/2011ApJ...726...95D}
}

@ARTICLE{allart_pastobs_2023,
       author = {{Allart}, R. and {Lem{\'e}e-Joliecoeur}, P. -B. and {Jaziri}, A.~Y. and {Lafreni{\`e}re}, D. and {Artigau}, E. and {Cook}, N. and {Darveau-Bernier}, A. and {Dang}, L. and {Cadieux}, C. and {Boucher}, A. and {Bourrier}, V. and {Deibert}, E.~K. and {Pelletier}, S. and {Radica}, M. and {Benneke}, B. and {Carmona}, A. and {Cloutier}, R. and {Cowan}, N.~B. and {Delfosse}, X. and {Donati}, J. -F. and {Doyon}, R. and {Figueira}, P. and {Forveille}, T. and {Fouqu{\'e}}, P. and {Gaidos}, E. and {Gu}, P. -G. and {H{\'e}brard}, G. and {Kiefer}, F. and {K{\'o}sp{\'a}l}, {\'A}. and {Jayawardhana}, R. and {Martioli}, E. and {Dos Santos}, L.~A. and {Shang}, H. and {Turner}, J.~D. and {Vidotto}, A.~A.},
        title = "{Homogeneous search for helium in the atmosphere of 11 gas giant exoplanets with SPIRou}",
      journal = {\aap},
         year = 2023,
        month = sep,
       volume = {677},
          eid = {A164},
        pages = {A164},
          doi = {10.1051/0004-6361/202245832},
archivePrefix = {arXiv},
       eprint = {2307.05580},
 primaryClass = {astro-ph.EP},
       adsurl = {https://ui.adsabs.harvard.edu/abs/2023A&A...677A.164A}
}

@ARTICLE{landman_pastobs_2024,
       author = {{Landman}, R. and {Stolker}, T. and {Snellen}, I.~A.~G. and {Costes}, J. and {de Regt}, S. and {Zhang}, Y. and {Gandhi}, S. and {Molliere}, P. and {Kesseli}, A. and {Vigan}, A. and {Sanchez-L{\'o}pez}, A.},
        title = "{{\ensuremath{\beta}} Pictoris b through the eyes of the upgraded CRIRES+. Atmospheric composition, spin rotation, and radial velocity}",
      journal = {\aap},
         year = 2024,
        month = feb,
       volume = {682},
          eid = {A48},
        pages = {A48},
          doi = {10.1051/0004-6361/202347846},
archivePrefix = {arXiv},
       eprint = {2311.13527},
 primaryClass = {astro-ph.EP},
       adsurl = {https://ui.adsabs.harvard.edu/abs/2024A&A...682A..48L}
}

@ARTICLE{smith_pastobs_2024,
       author = {{Smith}, Peter C.~B. and {Line}, Michael R. and {Bean}, Jacob L. and {Brogi}, Matteo and {August}, Prune and {Welbanks}, Luis and {Desert}, Jean-Michel and {Lunine}, Jonathan and {Sanchez}, Jorge and {Mansfield}, Megan and {Pino}, Lorenzo and {Rauscher}, Emily and {Kempton}, Eliza and {Zalesky}, Joseph and {Fowler}, Martin},
        title = "{A Combined Ground-based and JWST Atmospheric Retrieval Analysis: Both IGRINS and NIRSpec Agree that the Atmosphere of WASP-77A b Is Metal-poor}",
      journal = {\aj},
         year = 2024,
        month = mar,
       volume = {167},
       number = {3},
          eid = {110},
        pages = {110},
          doi = {10.3847/1538-3881/ad17bf},
archivePrefix = {arXiv},
       eprint = {2312.13069},
 primaryClass = {astro-ph.EP},
       adsurl = {https://ui.adsabs.harvard.edu/abs/2024AJ....167..110S}
}

@ARTICLE{brogi_pastobs_2023,
       author = {{Brogi}, Matteo and {Emeka-Okafor}, Vanessa and {Line}, Michael R. and {Gandhi}, Siddharth and {Pino}, Lorenzo and {Kempton}, Eliza M. -R. and {Rauscher}, Emily and {Parmentier}, Vivien and {Bean}, Jacob L. and {Mace}, Gregory N. and {Cowan}, Nicolas B. and {Shkolnik}, Evgenya and {Wardenier}, Joost P. and {Mansfield}, Megan and {Welbanks}, Luis and {Smith}, Peter and {Fortney}, Jonathan J. and {Birkby}, Jayne L. and {Zalesky}, Joseph A. and {Dang}, Lisa and {Patience}, Jennifer and {D{\'e}sert}, Jean-Michel},
        title = "{The Roasting Marshmallows Program with IGRINS on Gemini South I: Composition and Climate of the Ultrahot Jupiter WASP-18 b}",
      journal = {\aj},
         year = 2023,
        month = mar,
       volume = {165},
       number = {3},
          eid = {91},
        pages = {91},
          doi = {10.3847/1538-3881/acaf5c},
archivePrefix = {arXiv},
       eprint = {2209.15548},
 primaryClass = {astro-ph.EP},
       adsurl = {https://ui.adsabs.harvard.edu/abs/2023AJ....165...91B}
}

@inproceedings{chan_igrins_2014,
	author = {Chan Park and Daniel T. Jaffe and In-Soo Yuk and Moo-Young Chun and Soojong Pak and Kang-Min Kim and Michael Pavel and Hanshin Lee and Heeyoung Oh and Ueejeong Jeong and Chae Kyung Sim and Hye-In Lee and Huynh Anh Nguyen Le and Joseph Strubhar and Michael Gully-Santiago and Jae Sok Oh and Sang-Mok Cha and Bongkon Moon and Kwijong Park and Cynthia Brooks and Kyeongyeon Ko and Jeong-Yeol Han and Jakyoung Nah and Peter C. Hill and Sungho Lee and Stuart Barnes and Young Sam Yu and Kyle Kaplan and Gregory Mace and Hwihyun Kim and Jae-Joon Lee and Narae Hwang and Byeong-Gon Park},
	booktitle = {Ground-based and Airborne Instrumentation for Astronomy V},
	doi = {10.1117/12.2056431},
	editor = {Suzanne K. Ramsay and Ian S. McLean and Hideki Takami},
	organization = {International Society for Optics and Photonics},
	pages = {91471D},
	publisher = {SPIE},
	title = {{Design and early performance of IGRINS (Immersion Grating Infrared Spectrometer)}},
	url = {https://doi.org/10.1117/12.2056431},
	volume = {9147},
	year = {2014}}

@misc{lee_igpipe_2016,
       author = {{Lee}, Jae-Joon and {Gullikson}, Kevin},
        title = "{plp: v2.1 alpha 3}",
         year = 2016,
        month = jun,
          eid = {10.5281/zenodo.56067},
          doi = {10.5281/zenodo.56067},
      version = {v2.1-alpha.3},
    publisher = {Zenodo},
       adsurl = {https://ui.adsabs.harvard.edu/abs/2016zndo.....56067L}
}

@inproceedings{gregory_igpipe_2018,
	author = {Gregory Mace and Kimberly Sokal and Jae-Joon Lee and Heeyoung Oh and Chan Park and Hanshin Lee and John Good and Phillip MacQueen and Jae Sok Oh and Kyle Kaplan and Ben Kidder and Moo-Young Chun and In-Soo Yuk and Ueejeong Jeong and Soojong Pak and Kang-Min Kim and Jakyoung Nah and Sungho Lee and Young-Sam Yu and Narae Hwang and Byeong-Gon Park and Hwihyun Kim and Brian Chinn and Alison Peck and Ruben Diaz and Rene Rutten and Lisa Prato and George Jacoby and Frank Cornelius and Ben Hardesty and William DeGroff and Edward Dunham and Stephen Levine and Larissa Nofi and Ricardo Lopez-Valdivia and Alycia J. Weinberger and Daniel T. Jaffe},
	booktitle = {Ground-based and Airborne Instrumentation for Astronomy VII},
	doi = {10.1117/12.2312345},
	editor = {Christopher J. Evans and Luc Simard and Hideki Takami},
	organization = {International Society for Optics and Photonics},
	pages = {107020Q},
	publisher = {SPIE},
	title = {{IGRINS at the Discovery Channel Telescope and Gemini South}},
	url = {https://doi.org/10.1117/12.2312345},
	volume = {10702},
	year = {2018}}

@ARTICLE{molliere_prt_2020,
       author = {{Molli{\`e}re}, P. and {Stolker}, T. and {Lacour}, S. and {Otten}, G.~P.~P.~L. and {Shangguan}, J. and {Charnay}, B. and {Molyarova}, T. and {Nowak}, M. and {Henning}, Th. and {Marleau}, G. -D. and {Semenov}, D.~A. and {van Dishoeck}, E. and {Eisenhauer}, F. and {Garcia}, P. and {Garcia Lopez}, R. and {Girard}, J.~H. and {Greenbaum}, A.~Z. and {Hinkley}, S. and {Kervella}, P. and {Kreidberg}, L. and {Maire}, A. -L. and {Nasedkin}, E. and {Pueyo}, L. and {Snellen}, I.~A.~G. and {Vigan}, A. and {Wang}, J. and {de Zeeuw}, P.~T. and {Zurlo}, A.},
        title = "{Retrieving scattering clouds and disequilibrium chemistry in the atmosphere of HR 8799e}",
      journal = {\aap},
         year = 2020,
        month = aug,
       volume = {640},
          eid = {A131},
        pages = {A131},
          doi = {10.1051/0004-6361/202038325},
archivePrefix = {arXiv},
       eprint = {2006.09394},
 primaryClass = {astro-ph.EP},
       adsurl = {https://ui.adsabs.harvard.edu/abs/2020A&A...640A.131M}
}

@ARTICLE{molliere_prt_2019,
       author = {{Molli{\`e}re}, P. and {Wardenier}, J.~P. and {van Boekel}, R. and {Henning}, Th. and {Molaverdikhani}, K. and {Snellen}, I.~A.~G.},
        title = "{petitRADTRANS. A Python radiative transfer package for exoplanet characterization and retrieval}",
      journal = {\aap},
         year = 2019,
        month = jul,
       volume = {627},
          eid = {A67},
        pages = {A67},
          doi = {10.1051/0004-6361/201935470},
archivePrefix = {arXiv},
       eprint = {1904.11504},
 primaryClass = {astro-ph.EP},
       adsurl = {https://ui.adsabs.harvard.edu/abs/2019A&A...627A..67M}
}

@ARTICLE{guillot_prt_2010,
       author = {{Guillot}, T.},
        title = "{On the radiative equilibrium of irradiated planetary atmospheres}",
      journal = {\aap},
         year = 2010,
        month = sep,
       volume = {520},
          eid = {A27},
        pages = {A27},
          doi = {10.1051/0004-6361/200913396},
archivePrefix = {arXiv},
       eprint = {1006.4702},
 primaryClass = {astro-ph.EP},
       adsurl = {https://ui.adsabs.harvard.edu/abs/2010A&A...520A..27G}
}

@ARTICLE{gibson_equation_2020,
       author = {{Gibson}, Neale P. and {Merritt}, Stephanie and {Nugroho}, Stevanus K. and {Cubillos}, Patricio E. and {de Mooij}, Ernst J.~W. and {Mikal-Evans}, Thomas and {Fossati}, Luca and {Lothringer}, Joshua and {Nikolov}, Nikolay and {Sing}, David K. and {Spake}, Jessica J. and {Watson}, Chris A. and {Wilson}, Jamie},
        title = "{Detection of Fe I in the atmosphere of the ultra-hot Jupiter WASP-121b, and a new likelihood-based approach for Doppler-resolved spectroscopy}",
      journal = {\mnras},
         year = 2020,
        month = apr,
       volume = {493},
       number = {2},
        pages = {2215-2228},
          doi = {10.1093/mnras/staa228},
archivePrefix = {arXiv},
       eprint = {2001.06430},
 primaryClass = {astro-ph.EP},
       adsurl = {https://ui.adsabs.harvard.edu/abs/2020MNRAS.493.2215G}
}

@ARTICLE{brogi_equation_2019,
       author = {{Brogi}, Matteo and {Line}, Michael R.},
        title = "{Retrieving Temperatures and Abundances of Exoplanet Atmospheres with High-resolution Cross-correlation Spectroscopy}",
      journal = {\aj},
         year = 2019,
        month = mar,
       volume = {157},
       number = {3},
          eid = {114},
        pages = {114},
          doi = {10.3847/1538-3881/aaffd3},
archivePrefix = {arXiv},
       eprint = {1811.01681},
 primaryClass = {astro-ph.EP},
       adsurl = {https://ui.adsabs.harvard.edu/abs/2019AJ....157..114B}
}

@ARTICLE{foreman_emcee_2013,
       author = {{Foreman-Mackey}, Daniel and {Hogg}, David W. and {Lang}, Dustin and {Goodman}, Jonathan},
        title = "{emcee: The MCMC Hammer}",
      journal = {\pasp},
         year = 2013,
        month = mar,
       volume = {125},
       number = {925},
        pages = {306},
          doi = {10.1086/670067},
archivePrefix = {arXiv},
       eprint = {1202.3665},
 primaryClass = {astro-ph.IM},
       adsurl = {https://ui.adsabs.harvard.edu/abs/2013PASP..125..306F}
}

@article{virtanen_scipy_2020,
	title = {{SciPy} 1.0: fundamental algorithms for scientific computing in {Python}},
	volume = {17},
	copyright = {2020 The Author(s)},
	issn = {1548-7105},
	shorttitle = {{SciPy} 1.0},
	doi = {10.1038/s41592-019-0686-2},
	language = {en},
	number = {3},
	urldate = {2021-04-29},
	journal = {Nature Methods},
	author = {Virtanen, Pauli and Gommers, Ralf and Oliphant, Travis E. and Haberland, Matt and Reddy, Tyler and Cournapeau, David and Burovski, Evgeni and Peterson, Pearu and Weckesser, Warren and Bright, Jonathan and van der Walt, Stéfan J. and Brett, Matthew and Wilson, Joshua and Millman, K. Jarrod and Mayorov, Nikolay and Nelson, Andrew R. J. and Jones, Eric and Kern, Robert and Larson, Eric and Carey, C. J. and Polat, İlhan and Feng, Yu and Moore, Eric W. and VanderPlas, Jake and Laxalde, Denis and Perktold, Josef and Cimrman, Robert and Henriksen, Ian and Quintero, E. A. and Harris, Charles R. and Archibald, Anne M. and Ribeiro, Antônio H. and Pedregosa, Fabian and van Mulbregt, Paul},
	month = mar,
	year = {2020},
	pages = {261--272},
}

@ARTICLE{WeinerMansfield2024,
       author = {{Weiner Mansfield}, Megan and {Line}, Michael R. and {Wardenier}, Joost P. and {Brogi}, Matteo and {Bean}, Jacob L. and {Beltz}, Hayley and {Smith}, Peter and {Zalesky}, Joseph A. and {Batalha}, Natasha and {Kempton}, Eliza M. -R. and {Montet}, Benjamin T. and {Owen}, James E. and {Plavchan}, Peter and {Rauscher}, Emily},
        title = "{The Metallicity and Carbon-to-oxygen Ratio of the Ultrahot Jupiter WASP-76b from Gemini-S/IGRINS}",
      journal = {\aj},
         year = 2024,
        month = jul,
       volume = {168},
       number = {1},
          eid = {14},
        pages = {14},
          doi = {10.3847/1538-3881/ad4a5f},
archivePrefix = {arXiv},
       eprint = {2405.09769},
 primaryClass = {astro-ph.EP},
       adsurl = {https://ui.adsabs.harvard.edu/abs/2024AJ....168...14W}
}

@ARTICLE{Keating_2019,
       author = {{Keating}, Dylan and {Cowan}, Nicolas B. and {Dang}, Lisa},
        title = "{Uniformly hot nightside temperatures on short-period gas giants}",
      journal = {Nature Astronomy},
         year = 2019,
        month = aug,
       volume = {3},
        pages = {1092-1098},
          doi = {10.1038/s41550-019-0859-z},
archivePrefix = {arXiv},
       eprint = {1809.00002},
 primaryClass = {astro-ph.EP},
       adsurl = {https://ui.adsabs.harvard.edu/abs/2019NatAs...3.1092K}
}

@ARTICLE{Dang_2024,
       author = {{Dang}, Lisa and {Bell}, Taylor J. and {Ying} and {Shu} and {Cowan}, Nicolas B. and {Bean}, Jacob L. and {Deming}, Drake and {Kempton}, Eliza M. -R. and {Weiner Mansfield}, Megan and {Rauscher}, Emily and {Parmentier}, Vivien and {Stevenson}, Kevin B. and {Swain}, Mark and {Kreidberg}, Laura and {Kataria}, Tiffany and {D{\'e}sert}, Jean-Michel and {Zellem}, Robert and {Fortney}, Jonathan J. and {Lewis}, Nikole K. and {Line}, Michael and {Morley}, Caroline and {Showman}, Adam},
        title = "{A Comprehensive Analysis Spitzer 4.5 $\mu$m Phase Curve of Hot Jupiters}",
      journal = {arXiv e-prints},
         year = 2024,
        month = aug,
          eid = {arXiv:2408.13308},
        pages = {arXiv:2408.13308},
          doi = {10.48550/arXiv.2408.13308},
archivePrefix = {arXiv},
       eprint = {2408.13308},
 primaryClass = {astro-ph.EP},
       adsurl = {https://ui.adsabs.harvard.edu/abs/2024arXiv240813308D}
}

@INPROCEEDINGS{wong_2021,
       author = {{Wong}, Michael H. and {Luszcz-Cook}, Statia and {Sayanagi}, Kunio and {Moore}, Luke and {Koskinen}, Tommi and {Moses}, Julianne I. and {de Pater}, Imke and {Aslam}, Shahid and {Atreya}, Sushil K. and {Baines}, Kevin H. and {Bjoraker}, Gordon and {de Kleer}, Katherine R. and {Edgington}, Scott G. and {Fortney}, Jonathan and {Greathouse}, Thomas K. and {Hammel}, Heidi B. and {Li}, Cheng and {Mahaffy}, Paul R. and {Sinclair}, James and {Sromovsky}, Lawrence A.},
        title = "{Gas Giant and Ice Giant Atmospheres: Focused Questions for 2023-2032}",
    booktitle = {Bulletin of the American Astronomical Society},
         year = 2021,
       volume = {53},
        month = may,
          eid = {275},
        pages = {275},
          doi = {10.3847/25c2cfeb.f7906305},
       adsurl = {https://ui.adsabs.harvard.edu/abs/2021BAAS...53d.275W}
}

@ARTICLE{showman_2002,
       author = {{Showman}, A.~P. and {Guillot}, T.},
        title = "{Atmospheric circulation and tides of ``51 Pegasus b-like'' planets}",
      journal = {\aap},
         year = 2002,
        month = apr,
       volume = {385},
        pages = {166-180},
          doi = {10.1051/0004-6361:20020101},
archivePrefix = {arXiv},
       eprint = {astro-ph/0202236},
 primaryClass = {astro-ph},
       adsurl = {https://ui.adsabs.harvard.edu/abs/2002A&A...385..166S}
}

@ARTICLE{knutson_2007,
       author = {{Knutson}, Heather A. and {Charbonneau}, David and {Allen}, Lori E. and {Fortney}, Jonathan J. and {Agol}, Eric and {Cowan}, Nicolas B. and {Showman}, Adam P. and {Cooper}, Curtis S. and {Megeath}, S. Thomas},
        title = "{A map of the day-night contrast of the extrasolar planet HD 189733b}",
      journal = {\nat},
         year = 2007,
        month = may,
       volume = {447},
       number = {7141},
        pages = {183-186},
          doi = {10.1038/nature05782},
archivePrefix = {arXiv},
       eprint = {0705.0993},
 primaryClass = {astro-ph},
       adsurl = {https://ui.adsabs.harvard.edu/abs/2007Natur.447..183K}
}

@ARTICLE{hindle_2021,
       author = {{Hindle}, A.~W. and {Bushby}, P.~J. and {Rogers}, T.~M.},
        title = "{The Magnetic Mechanism for Hotspot Reversals in Hot Jupiter Atmospheres}",
      journal = {\apj},
         year = 2021,
        month = dec,
       volume = {922},
       number = {2},
          eid = {176},
        pages = {176},
          doi = {10.3847/1538-4357/ac0e2e},
archivePrefix = {arXiv},
       eprint = {2107.07515},
 primaryClass = {astro-ph.EP},
       adsurl = {https://ui.adsabs.harvard.edu/abs/2021ApJ...922..176H}
}

@ARTICLE{hindle_2019,
       author = {{Hindle}, A.~W. and {Bushby}, P.~J. and {Rogers}, T.~M.},
        title = "{Shallow-water Magnetohydrodynamics for Westward Hotspots on Hot Jupiters}",
      journal = {\apjl},
         year = 2019,
        month = feb,
       volume = {872},
       number = {2},
          eid = {L27},
        pages = {L27},
          doi = {10.3847/2041-8213/ab05dd},
archivePrefix = {arXiv},
       eprint = {1902.09683},
 primaryClass = {astro-ph.EP},
       adsurl = {https://ui.adsabs.harvard.edu/abs/2019ApJ...872L..27H}
}

@ARTICLE{teinturier_2024,
       author = {{Teinturier}, L. and {Charnay}, B. and {Spiga}, A. and {B{\'e}zard}, B. and {Leconte}, J. and {Mechineau}, A. and {Ducrot}, E. and {Millour}, E. and {Cl{\'e}ment}, N.},
        title = "{The radiative and dynamical impact of clouds in the atmosphere of the hot Jupiter WASP-43 b}",
      journal = {\aap},
         year = 2024,
        month = mar,
       volume = {683},
          eid = {A231},
        pages = {A231},
          doi = {10.1051/0004-6361/202347069},
archivePrefix = {arXiv},
       eprint = {2401.14083},
 primaryClass = {astro-ph.EP},
       adsurl = {https://ui.adsabs.harvard.edu/abs/2024A&A...683A.231T}
}

@ARTICLE{roman_2021,
       author = {{Roman}, Michael T. and {Kempton}, Eliza M. -R. and {Rauscher}, Emily and {Harada}, Caleb K. and {Bean}, Jacob L. and {Stevenson}, Kevin B.},
        title = "{Clouds in Three-dimensional Models of Hot Jupiters over a Wide Range of Temperatures. I. Thermal Structures and Broadband Phase-curve Predictions}",
      journal = {\apj},
         year = 2021,
        month = feb,
       volume = {908},
       number = {1},
          eid = {101},
        pages = {101},
          doi = {10.3847/1538-4357/abd549},
archivePrefix = {arXiv},
       eprint = {2010.06936},
 primaryClass = {astro-ph.EP},
       adsurl = {https://ui.adsabs.harvard.edu/abs/2021ApJ...908..101R}
}

@article{bitsch_2022,
	author = {{Bitsch, Bertram} and {Schneider, Aaron David} and {Kreidberg, Laura}},
	doi = {10.1051/0004-6361/202243345},
	journal = {A\&A},
	pages = {A138},
	title = {How drifting and evaporating pebbles shape giant planets - III. The formation of WASP-77A b and τ Bo{\"o}tis b},
	url = {https://doi.org/10.1051/0004-6361/202243345},
	volume = 665,
	year = 2022}

@article{coria_2024,
	author = {Coria, David R. and Hejazi, Neda and Crossfield, Ian J. M. and Rhem, Maleah},
	doi = {10.3847/1538-4357/ad7020},
	journal = {The Astrophysical Journal},
	month = {oct},
	number = {2},
	pages = {151},
	publisher = {The American Astronomical Society},
	title = {The Wanderer: Charting WASP-77A b's Formation and Migration Using a System-wide Inventory of Carbon and Oxygen Abundances},
	url = {https://dx.doi.org/10.3847/1538-4357/ad7020},
	volume = {974},
	year = {2024}}

@article{khorshid_2023,
	author = {{Khorshid, N.} and {Min, M.} and {D{\'e}sert, J. M.}},
	doi = {10.1051/0004-6361/202245469},
	journal = {A\&A},
	pages = {A95},
	title = {Retrieving planet formation parameters of WASP-77Ab using SimAb},
	url = {https://doi.org/10.1051/0004-6361/202245469},
	volume = 675,
	year = 2023}

@ARTICLE{Moses_2013,
       author = {{Moses}, J.~I. and {Madhusudhan}, N. and {Visscher}, C. and {Freedman}, R.~S.},
        title = "{Chemical Consequences of the C/O Ratio on Hot Jupiters: Examples from WASP-12b, CoRoT-2b, XO-1b, and HD 189733b}",
      journal = {\apj},
         year = 2013,
        month = jan,
       volume = {763},
       number = {1},
          eid = {25},
        pages = {25},
          doi = {10.1088/0004-637X/763/1/25},
archivePrefix = {arXiv},
       eprint = {1211.2996},
 primaryClass = {astro-ph.EP},
       adsurl = {https://ui.adsabs.harvard.edu/abs/2013ApJ...763...25M}
}

@ARTICLE{rauscher_2014,
       author = {{Rauscher}, Emily and {Kempton}, Eliza M.~R.},
        title = "{The Atmospheric Circulation and Observable Properties of Non-synchronously Rotating Hot Jupiters}",
      journal = {\apj},
         year = 2014,
        month = jul,
       volume = {790},
       number = {1},
          eid = {79},
        pages = {79},
          doi = {10.1088/0004-637X/790/1/79},
archivePrefix = {arXiv},
       eprint = {1402.4833},
 primaryClass = {astro-ph.EP},
       adsurl = {https://ui.adsabs.harvard.edu/abs/2014ApJ...790...79R}
}

@ARTICLE{pino_2022,
       author = {{Pino}, L. and {Brogi}, M. and {D{\'e}sert}, J.~M. and {Nascimbeni}, V. and {Bonomo}, A.~S. and {Rauscher}, E. and {Basilicata}, M. and {Biazzo}, K. and {Bignamini}, A. and {Borsa}, F. and {Claudi}, R. and {Covino}, E. and {Di Mauro}, M.~P. and {Guilluy}, G. and {Maggio}, A. and {Malavolta}, L. and {Micela}, G. and {Molinari}, E. and {Molinaro}, M. and {Montalto}, M. and {Nardiello}, D. and {Pedani}, M. and {Piotto}, G. and {Poretti}, E. and {Rainer}, M. and {Scandariato}, G. and {Sicilia}, D. and {Sozzetti}, A.},
        title = "{The GAPS Programme at TNG. XLI. The climate of KELT-9b revealed with a new approach to high-spectral-resolution phase curves}",
      journal = {\aap},
         year = 2022,
        month = dec,
       volume = {668},
          eid = {A176},
        pages = {A176},
          doi = {10.1051/0004-6361/202244593},
archivePrefix = {arXiv},
       eprint = {2209.11735},
 primaryClass = {astro-ph.EP},
       adsurl = {https://ui.adsabs.harvard.edu/abs/2022A&A...668A.176P}
}

@ARTICLE{wardenier_2024,
       author = {{Wardenier}, Joost P. and {Parmentier}, Vivien and {Line}, Michael R. and {Weiner Mansfield}, Megan and {Tan}, Xianyu and {Tsai}, Shang-Min and {Bean}, Jacob L. and {Birkby}, Jayne L. and {Brogi}, Matteo and {D{\'e}sert}, Jean-Michel and {Gandhi}, Siddharth and {Lee}, Elspeth K.~H. and {Levens}, Colette I. and {Pino}, Lorenzo and {Smith}, Peter C.~B.},
        title = "{Phase-resolving the Absorption Signatures of Water and Carbon Monoxide in the Atmosphere of the Ultra-hot Jupiter WASP-121b with GEMINI-S/IGRINS}",
      journal = {\pasp},
         year = 2024,
        month = aug,
       volume = {136},
       number = {8},
          eid = {084403},
        pages = {084403},
          doi = {10.1088/1538-3873/ad5c9f},
archivePrefix = {arXiv},
       eprint = {2406.09641},
 primaryClass = {astro-ph.EP},
       adsurl = {https://ui.adsabs.harvard.edu/abs/2024PASP..136h4403W}
}

@ARTICLE{van_sluijs_2023,
       author = {{van Sluijs}, Lennart and {Birkby}, Jayne L. and {Lothringer}, Joshua and {Lee}, Elspeth K.~H. and {Crossfield}, Ian J.~M. and {Parmentier}, Vivien and {Brogi}, Matteo and {Kulesa}, Craig and {McCarthy}, Don and {Charbonneau}, David},
        title = "{Carbon monoxide emission lines reveal an inverted atmosphere in the ultra hot Jupiter WASP-33 b consistent with an eastward hot spot}",
      journal = {\mnras},
         year = 2023,
        month = jun,
       volume = {522},
       number = {2},
        pages = {2145-2170},
          doi = {10.1093/mnras/stad1103},
archivePrefix = {arXiv},
       eprint = {2203.13234},
 primaryClass = {astro-ph.EP},
       adsurl = {https://ui.adsabs.harvard.edu/abs/2023MNRAS.522.2145V}
}

@ARTICLE{ejrenreich_2020,
       author = {{Ehrenreich}, David and {Lovis}, Christophe and {Allart}, Romain and {Zapatero Osorio}, Mar{\'\i}a Rosa and {Pepe}, Francesco and {Cristiani}, Stefano and {Rebolo}, Rafael and {Santos}, Nuno C. and {Borsa}, Francesco and {Demangeon}, Olivier and {Dumusque}, Xavier and {Gonz{\'a}lez Hern{\'a}ndez}, Jonay I. and {Casasayas-Barris}, N{\'u}ria and {S{\'e}gransan}, Damien and {Sousa}, S{\'e}rgio and {Abreu}, Manuel and {Adibekyan}, Vardan and {Affolter}, Michael and {Allende Prieto}, Carlos and {Alibert}, Yann and {Aliverti}, Matteo and {Alves}, David and {Amate}, Manuel and {Avila}, Gerardo and {Baldini}, Veronica and {Bandy}, Timothy and {Benz}, Willy and {Bianco}, Andrea and {Bolmont}, {\'E}meline and {Bouchy}, Fran{\c{c}}ois and {Bourrier}, Vincent and {Broeg}, Christopher and {Cabral}, Alexandre and {Calderone}, Giorgio and {Pall{\'e}}, Enric and {Cegla}, H.~M. and {Cirami}, Roberto and {Coelho}, Jo{\~a}o M.~P. and {Conconi}, Paolo and {Coretti}, Igor and {Cumani}, Claudio and {Cupani}, Guido and {Dekker}, Hans and {Delabre}, Bernard and {Deiries}, Sebastian and {D'Odorico}, Valentina and {Di Marcantonio}, Paolo and {Figueira}, Pedro and {Fragoso}, Ana and {Genolet}, Ludovic and {Genoni}, Matteo and {G{\'e}nova Santos}, Ricardo and {Hara}, Nathan and {Hughes}, Ian and {Iwert}, Olaf and {Kerber}, Florian and {Knudstrup}, Jens and {Landoni}, Marco and {Lavie}, Baptiste and {Lizon}, Jean-Louis and {Lendl}, Monika and {Lo Curto}, Gaspare and {Maire}, Charles and {Manescau}, Antonio and {Martins}, C.~J.~A.~P. and {M{\'e}gevand}, Denis and {Mehner}, Andrea and {Micela}, Giusi and {Modigliani}, Andrea and {Molaro}, Paolo and {Monteiro}, Manuel and {Monteiro}, Mario and {Moschetti}, Manuele and {M{\"u}ller}, Eric and {Nunes}, Nelson and {Oggioni}, Luca and {Oliveira}, Ant{\'o}nio and {Pariani}, Giorgio and {Pasquini}, Luca and {Poretti}, Ennio and {Rasilla}, Jos{\'e} Luis and {Redaelli}, Edoardo and {Riva}, Marco and {Santana Tschudi}, Samuel and {Santin}, Paolo and {Santos}, Pedro and {Segovia Milla}, Alex and {Seidel}, Julia V. and {Sosnowska}, Danuta and {Sozzetti}, Alessandro and {Span{\`o}}, Paolo and {Su{\'a}rez Mascare{\~n}o}, Alejandro and {Tabernero}, Hugo and {Tenegi}, Fabio and {Udry}, St{\'e}phane and {Zanutta}, Alessio and {Zerbi}, Filippo},
        title = "{Nightside condensation of iron in an ultrahot giant exoplanet}",
      journal = {\nat},
         year = 2020,
        month = apr,
       volume = {580},
       number = {7805},
        pages = {597-601},
          doi = {10.1038/s41586-020-2107-1},
archivePrefix = {arXiv},
       eprint = {2003.05528},
 primaryClass = {astro-ph.EP},
       adsurl = {https://ui.adsabs.harvard.edu/abs/2020Natur.580..597E}
}

@ARTICLE{bouchy_2017,
       author = {{Bouchy}, F. and {Doyon}, R. and {Artigau}, {\'E}. and {Melo}, C. and {Hernandez}, O. and {Wildi}, F. and {Delfosse}, X. and {Lovis}, C. and {Figueira}, P. and {Canto Martins}, B.~L. . and {Gonz{\'a}lez Hern{\'a}ndez}, J.~I. . and {Thibault}, S. and {Reshetov}, V. and {Pepe}, F. and {Santos}, N.~C. and {de Medeiros}, J.~R. . and {Rebolo}, R. and {Abreu}, M. and {Adibekyan}, V.~Z. and {Bandy}, T. and {Benz}, W. and {Blind}, N. and {Bohlender}, D. and {Boisse}, I. and {Bovay}, S. and {Broeg}, C. and {Brousseau}, D. and {Cabral}, A. and {Chazelas}, B. and {Cloutier}, R. and {Coelho}, J. and {Conod}, U. and {Cumming}, A. and {Delabre}, B. and {Genolet}, L. and {Hagelberg}, J. and {Jayawardhana}, R. and {K{\"a}ufl}, H. -U. and {Lafreni{\`e}re}, D. and {de Castro Le{\~a}o}, I. . and {Malo}, L. and {de Medeiros Martins}, A. . and {Matthews}, J.~M. and {Metchev}, S. and {Oshagh}, M. and {Ouellet}, M. and {Parro}, V.~C. and {Rasilla Pi{\~n}eiro}, J.~L. . and {Santos}, P. and {Sarajlic}, M. and {Segovia}, A. and {Sordet}, M. and {Udry}, S. and {Valencia}, D. and {Vall{\'e}e}, P. and {Venn}, K. and {Wade}, G.~A. and {Saddlemyer}, L.},
        title = "{Near-InfraRed Planet Searcher to Join HARPS on the ESO 3.6-metre Telescope}",
      journal = {The Messenger},
         year = 2017,
        month = sep,
       volume = {169},
        pages = {21-27},
          doi = {10.18727/0722-6691/5034},
       adsurl = {https://ui.adsabs.harvard.edu/abs/2017Msngr.169...21B}
}

@ARTICLE{artigau_2024,
       author = {{Artigau}, {\'E}tienne and {Bouchy}, Fran{\c{c}}ois and {Doyon}, Ren{\'e} and {Baron}, Fr{\'e}d{\'e}rique and {Malo}, Lison and {Wildi}, Fran{\c{c}}ois and {Pepe}, Franceso and {Cook}, Neil J. and {Thibault}, Simon and {Reshetov}, Vladimir and {Dumusque}, Xavier and {Lovis}, Christophe and {Sosnowska}, Danuta and {Canto Martins}, Bruno L. and {Renan De Medeiros}, Jose and {Delfosse}, Xavier and {Santos}, Nuno and {Rebolo}, Rafael and {Abreu}, Manuel and {Allain}, Guillaume and {Allart}, Romain and {Auger}, Hugues and {Barros}, Susana and {Bazinet}, Luc and {Blind}, Nicolas and {Boisse}, Isabelle and {Bonfils}, Xavier and {Bourrier}, Vincent and {Bovay}, S{\'e}bastien and {Broeg}, Christopher and {Brousseau}, Denis and {Bruniquel}, Vincent and {Cabral}, Alexandre and {Cadieux}, Charles and {Carmona}, Andres and {Carteret}, Yann and {Challita}, Zalpha and {Chazelas}, Bruno and {Cloutier}, Ryan and {Coelho}, Jo{\~a}o and {Cointepas}, Marion and {Conod}, Uriel and {Cowan}, Nicolas and {Cristo}, Eduardo and {Gomes da Silva}, Jo{\~a}o and {Dauplaise}, Laurie and {de Lima Gomes}, Roseane and {Delgado-Mena}, Elisa and {Ehrenreich}, David and {Faria}, Jo{\~a}o and {Figueira}, Pedro and {Forveille}, Thierry and {Frensch}, Yolanda and {Gagn{\'e}}, Jonathan and {Genest}, Fr{\'e}d{\'e}ric and {Genolet}, Ludovic and {Gonz{\'a}lez Hern{\'a}ndez}, Jonay I. and {T{\'e}mich}, F{\'e}lix Gracia and {Grieves}, Nolan and {Hernandez}, Olivier and {Hobson}, Melissa J. and {Hoeijmakers}, Jens and {Kerley}, Dan and {Krishnamurthy}, Vigneshwaran and {Lafreni{\`e}re}, David and {Lamontagne}, Pierrot and {Larue}, Pierre and {Leaf}, Henry and {Le{\~a}o}, Izan C. and {Lim}, Olivia and {Lo Curto}, Gaspare and {Martins}, Allan M. and {Melo}, Claudio and {Messias}, Yuri S. and {Mignon}, Lucile and {Moranta}, Leslie and {Mordasini}, Christoph and {Al Moulla}, Khaled and {Mounzer}, Dany and {L'Heureux}, Alexandrine and {Nari}, Nicola and {Nielsen}, Louise and {Osborn}, Ares and {Parc}, L{\'e}na and {Pasquini}, Luca and {Passegger}, Vera M. and {Pelletier}, Stefan and {Peroux}, C{\'e}line and {Piaulet}, Caroline and {Plotnykov}, Mykhaylo and {Poulin-Girard}, Anne-Sophie and {Rasilla}, Jos{\'e} Luis and {Saint-Antoine}, Jonathan and {Sarajlic}, Mirsad and {Segovia}, Alex and {Seidel}, Julia and {S{\'e}gransan}, Damien and {Costa Silva}, Ana Rita and {Srivastava}, Avidaan and {Stefanov}, Atanas K. and {Su{\'a}rez Mascare{\~n}o}, Alejandro and {Sordet}, Michael and {Teixeira}, M{\'a}rcio A. and {Udry}, St{\'e}phane and {Valencia}, Diana and {Vall{\'e}e}, Philippe and {Vandal}, Thomas and {Vaulato}, Valentina and {Wade}, Gregg and {Wardenier}, Joost P. and {Wehb{\'e}}, Bachar and {Weisserman}, Drew and {Wevers}, Ivan and {Zins}, G{\'e}rard},
        title = "{NIRPS first light and early science: breaking the 1 m/s RV precision barrier at infrared wavelengths}",
      journal = {arXiv e-prints},
         year = 2024,
        month = jun,
          eid = {arXiv:2406.08304},
        pages = {arXiv:2406.08304},
          doi = {10.48550/arXiv.2406.08304},
archivePrefix = {arXiv},
       eprint = {2406.08304},
 primaryClass = {astro-ph.IM},
       adsurl = {https://ui.adsabs.harvard.edu/abs/2024arXiv240608304A}
}

@ARTICLE{borsa_2021,
       author = {{Borsa}, Francesco and {Fossati}, Luca and {Koskinen}, Tommi and {Young}, Mitchell E. and {Shulyak}, Denis},
        title = "{High-resolution detection of neutral oxygen and non-LTE effects in the atmosphere of KELT-9b}",
      journal = {Nature Astronomy},
         year = 2021,
        month = dec,
       volume = {6},
        pages = {226-231},
          doi = {10.1038/s41550-021-01544-4},
archivePrefix = {arXiv},
       eprint = {2112.12059},
 primaryClass = {astro-ph.EP},
       adsurl = {https://ui.adsabs.harvard.edu/abs/2022NatAs...6..226B}
}

@article{Pelletier_2023,
       author = {Pelletier, Stefan and {Benneke}, Bj{\"o}rn and {Ali-Dib}, Mohamad and {Prinoth}, Bibiana and {Kasper}, David and {Seifahrt}, Andreas and {Bean}, Jacob L. and {Debras}, Florian and {Klein}, Baptiste and {Bazinet}, Luc and {Hoeijmakers}, H. Jens and {Kesseli}, Aurora Y. and {Lim}, Olivia and {Carmona}, Andres and {Pino}, Lorenzo and {Casasayas-Barris}, N{\'u}ria and {Hood}, Thea and {St{\"u}rmer}, Julian},
        title = "{Vanadium oxide and a sharp onset of cold-trapping on a giant exoplanet}",
      journal = {\nat},
         year = 2023,
        month = jul,
       volume = {619},
       number = {7970},
        pages = {491-494},
          doi = {10.1038/s41586-023-06134-0},
archivePrefix = {arXiv},
       eprint = {2306.08739},
 primaryClass = {astro-ph.EP},
       adsurl = {https://ui.adsabs.harvard.edu/abs/2023Natur.619..491P}
}

@article{pelletier_crires_2024,
	title = {{CRIRES}+ and {ESPRESSO} {Reveal} an {Atmosphere} {Enriched} in {Volatiles} {Relative} to {Refractories} on the {Ultrahot} {Jupiter} {WASP}-121b},
	volume = {169},
	issn = {1538-3881},
	url = {https://dx.doi.org/10.3847/1538-3881/ad8b28},
	doi = {10.3847/1538-3881/ad8b28},
	language = {en},
	number = {1},
	urldate = {2024-12-07},
	journal = {The Astronomical Journal},
	author = {Pelletier, Stefan and Benneke, Björn and Chachan, Yayaati and Bazinet, Luc and Allart, Romain and Hoeijmakers, H. Jens and Lavail, Alexis and Prinoth, Bibiana and Coulombe, Louis-Philippe and Lothringer, Joshua D. and Parmentier, Vivien and Smith, Peter and Borsato, Nicholas and Thorsbro, Brian},
	month = dec,
	year = {2024},
	pages = {10},
}

@INPROCEEDINGS{oliva_2018,
       author = {{Oliva}, E. and {Sanna}, N. and {Rainer}, M. and {Massi}, F. and {Tozzi}, A. and {Origlia}, L.},
        title = "{GIANO, the high resolution IR spectrograph of the TNG: geometry of the echellogram and strategies for the 2D reduction of the spectra}",
    booktitle = {Ground-based and Airborne Instrumentation for Astronomy VII},
         year = 2018,
       editor = {{Evans}, Christopher J. and {Simard}, Luc and {Takami}, Hideki},
       series = {Society of Photo-Optical Instrumentation Engineers (SPIE) Conference Series},
       volume = {10702},
        month = jul,
          eid = {1070274},
        pages = {1070274},
          doi = {10.1117/12.2309927},
       adsurl = {https://ui.adsabs.harvard.edu/abs/2018SPIE10702E..74O}
}

@ARTICLE{czesla_2024,
       author = {{Czesla}, S. and {Lamp{\'o}n}, M. and {Cont}, D. and {Lesjak}, F. and {Orell-Miquel}, J. and {Sanz-Forcada}, J. and {Nagel}, E. and {Nortmann}, L. and {Molaverdikhani}, K. and {L{\'o}pez-Puertas}, M. and {Yan}, F. and {Quirrenbach}, A. and {Caballero}, J.~A. and {Pall{\'e}}, E. and {Aceituno}, J. and {Amado}, P.~J. and {Henning}, Th. and {Khalafinejad}, S. and {Montes}, D. and {Reiners}, A. and {Ribas}, I. and {Schweitzer}, A.},
        title = "{The elusive atmosphere of WASP-12 b. High-resolution transmission spectroscopy with CARMENES}",
      journal = {\aap},
         year = 2024,
        month = mar,
       volume = {683},
          eid = {A67},
        pages = {A67},
          doi = {10.1051/0004-6361/202348107},
archivePrefix = {arXiv},
       eprint = {2401.02195},
 primaryClass = {astro-ph.EP},
       adsurl = {https://ui.adsabs.harvard.edu/abs/2024A&A...683A..67C}
}

@INPROCEEDINGS{jones_2022,
       author = {{Jones}, Kathryn and {Morris}, Brett and {Demory}, Brice-Olivier and {Heng}, Kevin and {Hooton}, Matthew},
        title = "{On the climate of exoplanet KELT-9b with precision CHEOPS phase curves}",
    booktitle = {Bulletin of the American Astronomical Society},
         year = 2022,
       volume = {54},
        month = jun,
          eid = {102.128},
        pages = {102.128},
       adsurl = {https://ui.adsabs.harvard.edu/abs/2022BAAS...54e.128J}
}

@article{Alonso_2019,
	author = {{Alonso-Floriano, F. J.} and {S{\'a}nchez-L{\'o}pez, A.} and {Snellen, I. A. G.} and {L{\'o}pez-Puertas, M.} and {Nagel, E.} and {Amado, P. J.} and {Bauer, F. F.} and {Caballero, J. A.} and {Czesla, S.} and {Nortmann, L.} and {Pall{\'e}, E.} and {Salz, M.} and {Reiners, A.} and {Ribas, I.} and {Quirrenbach, A.} and {Aceituno, J.} and {Anglada-Escud{\'e}, G.} and {B{\'e}jar, V. J. S.} and {Guenther, E. W.} and {Henning, T.} and {Kaminski, A.} and {K{\"u}rster, M.} and {Lamp{\'o}n, M.} and {Lara, L. M.} and {Montes, D.} and {Morales, J. C.} and {Tal-Or, L.} and {Schmitt, J. H. M. M.} and {Zapatero Osorio, M. R.} and {Zechmeister, M.}},
	doi = {10.1051/0004-6361/201834339},
	journal = {A\&A},
	pages = {A74},
	title = {Multiple water band detections in the CARMENES near-infrared transmission spectrum of HD 189733 b},
	url = {https://doi.org/10.1051/0004-6361/201834339},
	volume = 621,
	year = 2019}

@article{Giacobbe_2021,
   author = {Giacobbe, Paolo and Brogi, Matteo and Gandhi, Siddharth and Cubillos, Patricio E. and Bonomo, Aldo S. and Sozzetti, Alessandro and Fossati, Luca and Guilluy, Gloria and Carleo, Ilaria and Rainer, Monica and Harutyunyan, Avet and Borsa, Francesco and Pino, Lorenzo and Nascimbeni, Valerio and Benatti, Serena and Biazzo, Katia and Bignamini, Andrea and Chubb, Katy L. and Claudi, Riccardo and Cosentino, Rosario and Covino, Elvira and Damasso, Mario and Desidera, Silvano and Fiorenzano, Aldo F. M. and Ghedina, Adriano and Lanza, Antonino F. and Leto, Giuseppe and Maggio, Antonio and Malavolta, Luca and Maldonado, Jesus and Micela, Giuseppina and Molinari, Emilio and Pagano, Isabella and Pedani, Marco and Piotto, Giampaolo and Poretti, Ennio and Scandariato, Gaetano and Yurchenko, Sergei N. and Fantinel, Daniela and Galli, Alberto and Lodi, Marcello and Sanna, Nicoletta and Tozzi, Andrea},
   title = {Five carbon- and nitrogen-bearing species in a hot giant planet’s atmosphere},
   journal = {Nature},
   volume = {592},
   number = {7853},
   pages = {205-208},
   ISSN = {1476-4687},
   DOI = {10.1038/s41586-021-03381-x},
   url = {https://doi.org/10.1038/s41586-021-03381-x},
   year = {2021},
   type = {Journal Article}
}

@article{carleo_2022,
  title={The GAPS Programme at TNG XXXIX. Multiple Molecular Species in the Atmosphere of the Warm Giant Planet WASP-80 b Unveiled at High Resolution with GIANO-B∗},
  author={Carleo, Ilaria and Giacobbe, Paolo and Guilluy, Gloria and Cubillos, Patricio E and Bonomo, Aldo S and Sozzetti, Alessandro and Brogi, Matteo and Gandhi, Siddharth and Fossati, Luca and Turrini, Diego and others},
  journal={The Astronomical Journal},
  volume={164},
  number={3},
  pages={101},
  year={2022},
  publisher={IOP Publishing}
}

@article{Guilluy_2022,
	author = {{Guilluy, G.} and {Giacobbe, P.} and {Carleo, I.} and {Cubillos, P. E.} and {Sozzetti, A.} and {Bonomo, A. S.} and {Brogi, M.} and {Gandhi, S.} and {Fossati, L.} and {Nascimbeni, V.} and {Turrini, D.} and {Schisano, E.} and {Borsa, F.} and {Lanza, A. F.} and {Mancini, L.} and {Maggio, A.} and {Malavolta, L.} and {Micela, G.} and {Pino, L.} and {Rainer, M.} and {Bignamini, A.} and {Claudi, R.} and {Cosentino, R.} and {Covino, E.} and {Desidera, S.} and {Fiorenzano, A.} and {Harutyunyan, A.} and {Lorenzi, V.} and {Knapic, C.} and {Molinari, E.} and {Pacetti, E.} and {Pagano, I.} and {Pedani, M.} and {Piotto, G.} and {Poretti, E.}},
	doi = {10.1051/0004-6361/202243854},
	journal = {A\&A},
	pages = {A104},
	title = {The GAPS Programme at TNG - XXXVIII. Five molecules in the atmosphere of the warm giant planet WASP-69b detected at high spectral resolution★},
	url = {https://doi.org/10.1051/0004-6361/202243854},
	volume = 665,
	year = 2022}

@article{benneke_atmospheric_2012,
	title = {{ATMOSPHERIC} {RETRIEVAL} {FOR} {SUPER}-{EARTHS}: {UNIQUELY} {CONSTRAINING} {THE} {ATMOSPHERIC} {COMPOSITION} {WITH} {TRANSMISSION} {SPECTROSCOPY}},
	volume = {753},
	issn = {0004-637X},
	shorttitle = {{ATMOSPHERIC} {RETRIEVAL} {FOR} {SUPER}-{EARTHS}},
	url = {https://dx.doi.org/10.1088/0004-637X/753/2/100},
	doi = {10.1088/0004-637X/753/2/100},
	language = {en},
	number = {2},
	urldate = {2023-04-19},
	journal = {The Astrophysical Journal},
	author = {Benneke, Bjoern and Seager, Sara},
	month = jun,
	year = {2012},
	pages = {100},
}

@article{benneke_how_2013,
	title = {{HOW} {TO} {DISTINGUISH} {BETWEEN} {CLOUDY} {MINI}-{NEPTUNES} {AND} {WATER}/{VOLATILE}-{DOMINATED} {SUPER}-{EARTHS}},
	volume = {778},
	issn = {0004-637X},
	url = {https://dx.doi.org/10.1088/0004-637X/778/2/153},
	doi = {10.1088/0004-637X/778/2/153},
	language = {en},
	number = {2},
	urldate = {2023-04-19},
	journal = {The Astrophysical Journal},
	author = {Benneke, Björn and Seager, Sara},
	month = nov,
	year = {2013},
	pages = {153},
}

@misc{benneke_strict_2015,
	title = {Strict {Upper} {Limits} on the {Carbon}-to-{Oxygen} {Ratios} of {Eight} {Hot} {Jupiters} from {Self}-{Consistent} {Atmospheric} {Retrieval}},
	url = {http://arxiv.org/abs/1504.07655},
	urldate = {2023-04-19},
	publisher = {arXiv},
	author = {Benneke, Björn},
	month = apr,
	year = {2015},
	doi = {10.48550/arXiv.1504.07655},
}

@ARTICLE{Chubb_2021,
       author = {{Chubb}, Katy L. and {Rocchetto}, Marco and {Yurchenko}, Sergei N. and {Min}, Michiel and {Waldmann}, Ingo and {Barstow}, Joanna K. and {Molli{\`e}re}, Paul and {Al-Refaie}, Ahmed F. and {Phillips}, Mark W. and {Tennyson}, Jonathan},
        title = "{The ExoMolOP database: Cross sections and k-tables for molecules of interest in high-temperature exoplanet atmospheres}",
      journal = {\aap},
         year = 2021,
        month = feb,
       volume = {646},
          eid = {A21},
        pages = {A21},
          doi = {10.1051/0004-6361/202038350},
archivePrefix = {arXiv},
       eprint = {2009.00687},
 primaryClass = {astro-ph.EP},
       adsurl = {https://ui.adsabs.harvard.edu/abs/2021A&A...646A..21C}
}

@article{Brooke_2016,
	author = {James S.A. Brooke and Peter F. Bernath and Colin M. Western and Christopher Sneden and Melike Af{\c s}ar and Gang Li and Iouli E. Gordon},
	doi = {https://doi.org/10.1016/j.jqsrt.2015.07.021},
	issn = {0022-4073},
	journal = {Journal of Quantitative Spectroscopy and Radiative Transfer},
	pages = {142-157},
	title = {Line strengths of rovibrational and rotational transitions in the X2Π ground state of OH},
	url = {https://www.sciencedirect.com/science/article/pii/S0022407315002721},
	volume = {168},
	year = {2016}}

@article{Yurchenko_2020,
   author = {Yurchenko, S N and Mellor, Thomas M and Freedman, Richard S and Tennyson, J},
   title = {ExoMol line lists – XXXIX. Ro-vibrational molecular line list for CO2},
   journal = {Monthly Notices of the Royal Astronomical Society},
   volume = {496},
   number = {4},
   pages = {5282-5291},
   ISSN = {0035-8711},
   DOI = {10.1093/mnras/staa1874},
   url = {https://doi.org/10.1093/mnras/staa1874},
   year = {2020},
   type = {Journal Article}
}

@ARTICLE{Hargreaves_2020,
       author = {{Hargreaves}, Robert J. and {Gordon}, Iouli E. and {Rey}, Michael and {Nikitin}, Andrei V. and {Tyuterev}, Vladimir G. and {Kochanov}, Roman V. and {Rothman}, Laurence S.},
        title = "{An Accurate, Extensive, and Practical Line List of Methane for the HITEMP Database}",
      journal = {\apjs},
         year = 2020,
        month = apr,
       volume = {247},
       number = {2},
          eid = {55},
        pages = {55},
          doi = {10.3847/1538-4365/ab7a1a},
archivePrefix = {arXiv},
       eprint = {2001.05037},
 primaryClass = {astro-ph.EP},
       adsurl = {https://ui.adsabs.harvard.edu/abs/2020ApJS..247...55H}
}

@article{McKemmish_2016,
   author = {McKemmish, Laura K. and Yurchenko, Sergei N. and Tennyson, Jonathan},
   title = {ExoMol line lists – XVIII. The high-temperature spectrum of VO},
   journal = {Monthly Notices of the Royal Astronomical Society},
   volume = {463},
   number = {1},
   pages = {771-793},
   ISSN = {0035-8711},
   DOI = {10.1093/mnras/stw1969},
   url = {https://doi.org/10.1093/mnras/stw1969},
   year = {2016},
   type = {Journal Article}
}

@book{burnham2002model,
  title={Model selection and multimodel inference: a practical information-theoretic approach},
  author={Burnham, Kenneth P and Anderson, David R},
  year={2002},
  publisher={Springer}
}

@ARTICLE{plez_tio,
       author = {{Plez}, B.},
        title = "{A new TiO line list}",
      journal = {\aap},
         year = 1998,
        month = sep,
       volume = {337},
        pages = {495-500},
       adsurl = {https://ui.adsabs.harvard.edu/abs/1998A&A...337..495P}
}

@ARTICLE{Rothman_hitemp,
       author = {{Rothman}, L.~S. and {Gordon}, I.~E. and {Barber}, R.~J. and {Dothe}, H. and {Gamache}, R.~R. and {Goldman}, A. and {Perevalov}, V.~I. and {Tashkun}, S.~A. and {Tennyson}, J.},
        title = "{HITEMP, the high-temperature molecular spectroscopic database}",
      journal = {\jqsrt},
         year = 2010,
        month = oct,
       volume = {111},
        pages = {2139-2150},
          doi = {10.1016/j.jqsrt.2010.05.001},
       adsurl = {https://ui.adsabs.harvard.edu/abs/2010JQSRT.111.2139R}
}

@ARTICLE{Barber_exomol,
       author = {{Barber}, R.~J. and {Strange}, J.~K. and {Hill}, C. and {Polyansky}, O.~L. and {Mellau}, G. Ch. and {Yurchenko}, S.~N. and {Tennyson}, Jonathan},
        title = "{ExoMol line lists - III. An improved hot rotation-vibration line list for HCN and HNC}",
      journal = {\mnras},
         year = 2014,
        month = jan,
       volume = {437},
       number = {2},
        pages = {1828-1835},
          doi = {10.1093/mnras/stt2011},
archivePrefix = {arXiv},
       eprint = {1311.1328},
 primaryClass = {astro-ph.SR},
       adsurl = {https://ui.adsabs.harvard.edu/abs/2014MNRAS.437.1828B}
}
\bibliographystyle{aasjournal}

\end{document}